\documentclass{jltp}

\title{Zurek-Kibble Causality Bounds in Time-Dependent Ginzburg-Landau Theory and Quantum Field Theory}

\author{R.J.~Rivers}

\address{Theoretical Physics, Blackett Laboratory, Imperial College, Prince\\
Consort Road, London, SW7 2BZ, U.K.}

\runninghead{R.J.~Rivers}{ZK Causality in TDGL Theory and QFT}

\begin{document}

\maketitle

\begin{abstract}
Zurek's and Kibble's causal constraints for defect production at continuous
transitions are encoded in the field equations that condensed matter systems and
quantum fields satisfy. In this article we
highlight some of the properties of the solutions to the equations
and show to what extent they support the original ideas.

\end{abstract}

\section{INTRODUCTION}

There is a tradition of condensed matter physicists and quantum
field theorists (particle physicists) exploring ideas and
techniques in common. The development of {\it equilibrium}
renormalisation group methods in the '70s and onwards by both
communities to describe phase transitions has been one of the
major successes of that period. More recently, the adoption of
simple causal bounds to constrain the {\it non-equilibrium}
dynamics of continuous phase transitions has led to a renewal of
this dialogue, reflected in the title {\it Cosmological
Experiments in Condensed Matter Systems} of the review article by
Zurek\cite{zurek1}, the first extensive presentation of these
ideas.

Although subsequent work by several authors (including Zurek
himself\cite{zurek2,zurek3}) has refined these original
proposals\cite{zurek2a} they are, at heart, very simple: after a
continuous transition  the order parameter field (or fields)
cannot adapt to a single uniform ground state value immediately.
 The reason is straightforward. Although the adiabatic (equilibrium)
correlation length $\xi_{eq}(t)$ diverges at the transition the
true correlation length $\xi (t)$ does not, since there is not
enough time for it to do so. Causality imposes a maximum rate at
which the correlation length can grow and hence a maximum
correlation length, ${\bar\xi}$ say, at the onset of the
transition. At the same time, causality imposes a horizon outside
which the fields are uncorrelated.
A consequence of a correlation length that is always
finite is the creation of topological defects (monopoles,
vortices, walls, etc.) that mark 'domain' boundaries. These
defects then self-interact and annihilate, leading to larger and
larger regions over which the field assumes one of the possible
ground state values.

If the simple relationship ${\bar\xi}= O({\bar\xi}_{def})$ between
the correlation length ${\bar\xi}$ and the separation length
${\bar\xi}_{def}$ of defects suggested by this picture is valid
then the density of defects is bounded, and calculable, at their
moment of formation. If their evolution is known, the density of
defects at late times is similarly constrained.

Since causality is as equally embodied in quantum field theory
(QFT) as in condensed matter, these ideas had been posed
independently by Kibble\cite{kibble1,kibble2} in the context of
phase transitions of QFT in the very early universe. One of the
great successes of QFT has been the unification of the electroweak
forces through spontaneous symmetry breaking. There is every
reason to believe that this and other symmetries were not always
broken but that, in the very early and hot universe, they were
restored. Kibble\cite{kibble3,kibble4} and others\cite{joao} have
observed that the same causal arguments put useful constraints on
the density of defects at the time of their formation, which could
have consequences today. Unfortunately, because of our lack of
understanding of the details of the early universe it is
impossible to make reliable predictions\cite{wein}.

It was Zurek who argued that the same causal bounds could be
tested directly in condensed matter systems, whose phase
transitions produce defects whose densities can be measured fairly
readily. In the final section of this paper we review most of the
experimental evidence for Zurek's bounds which, on several
occasions\cite{helsinki,grenoble,KMR}, are satisfied at a good
qualitative level.

Prior to that, in the next two sections we reiterate and rephrase
the Zurek-Kibble bounds before providing an alternative scenario
for the way defects appear after a transition. This is based on
the fact that the simple defects that we shall consider have
'false' ground-state or vacuum at their cores where the field {\it
vanishes}. Under suitable circumstances the separation length
${\bar\xi}_{def}$ is more sensibly derived by counting {\it
zeroes} of the field as ${\bar\xi}_{def}= O({\bar\xi}_{zero})$,
where ${\bar\xi}_{zero}$ measures the separation of field zeroes.
However, in principle, ${\bar\xi}_{zero}$ and ${\bar\xi}$ are {\it
different} correlation lengths, exploring different attributes of
the fluctuation spectrum.

We then apply these ideas in turn to condensed matter systems,
using the empirical time-dependent Ginzburg Landau (TDGL) theory
as specifying the dynamics, and to relativistic Quantum Field
Theory (QFT). In the latter case, there is a further complication
in that the system needs to decohere (i.e. lose its quantum
mechanical interference) before we can use classical probabilities
to identify defects, and we devote Section 6  to that.

Nonetheless, subject to certain conditions, we find that
${\bar\xi}_{def}$ is, indeed, $O({\bar\xi})$, as predicted, both
for condensed matter and QFT. The Kibble-Zurek bounds are
qualitatively valid, albeit for somewhat different reasons than
simple causality. A byproduct is that the failure of $^4 He$
experiments to give agreement\cite{lancaster2} may lie in an
incorrect assumption of the decay rate of defects.

We have not attempted to be exhaustive in this article and several
important topics have not been included (e.g. temperature
inhomogeneities). The reader is referred to the references for
more details.

\section{CAUSAL CONSTRAINTS ON THE PRODUCTION OF DEFECTS}

\subsection{Condensed Matter}

The role of defects in our preliminary discussion was to both to
infer and to predict the finiteness of the correlation length
after the transition. Both condensed matter systems ($^3 He$ in
particular\cite{volovik}) and QFT\cite{shellard} permit a
prodigious variety of defects, and we shall report on some of them
later. Rather than try to be generic we use the fact that most
(but not all) experiments involve vortices and, with this in mind,
we take an idealised superfluid as the prototypical condensed
matter system in which to display these arguments. An essential
property of superfluids is that their symmetries are {\it global}.
For simplicity, we assume a single complex order parameter field
$\phi ({\bf x},t)$, which permits the most simple
vortices\cite{Kleinert}.

Most simply, the transition from normal fluid to superfluid is
achieved through a pressure or temperature quench.
 The rate at which $\phi ({\bf x},t)$ can adjust, and the field become correlated, is
the speed of (second) sound ${\bar c}(\epsilon)$, where $\epsilon
= (T-T_c )/T_c$, which slows critically to zero as
$\epsilon\rightarrow 0$. If we know the time dependence $\epsilon
(t)$ of $(T-T_c )/T_c$ in the vicinity of the transition, the
diameter of the 'sonic horizon' at time $t$ is
\begin{equation}
h(t) = 2\int^{t}_{0}\,dt'\,{\bar c}(t'),
 \label{sonic}
\end{equation}
where ${\bar c}(t) = {\bar c}(\epsilon (t))$, and we have assumed
that the transition began at time $t=0$, for which $\epsilon (0) =
0$.

This immediately leads to a 'weak' causal constraint that, at time
$t$ after the transition has been completed, the correlation
length $\xi (t)$  satisfies the horizon bound
\begin{equation}
\xi (t) < h(t). \label{causalw}
\end{equation}
i.e. the field is uncorrelated outside its causal horizon. For
long times (\ref{causalw}) becomes $\xi (t)\leq 2{\bar
c}(\epsilon_f)$, where $\epsilon_f$ is the final relative
temperature.

At his stage we need to be more explicit in that, for our
prototypical case of a single complex field, there are two
candidate correlation lengths, one for the field magnitude (e.g.
superfluid density) and one for the phase correlation (e.g.
superflow). Once the transition is complete it is the latter,
associated with Goldstone modes, that determines the horizon.

However, for the purpose of experiment the more relevant bound
invoked by Zurek (and by Kibble in Ref.\onlinecite{kibble2}) is
the following 'strong' causal bound, for which this delineation is
not necessary, in the first instance: Zurek argued that the
correlation length freezes in {\it before} the transition, at $t =
-{\bar t}$, when its growth rate reaches its causal bound
${\dot\xi}_{eq} (-{\bar t})= +{\bar c}(-{\bar t})$, and only
unfreezes at a comparable time $+{\bar t}$ {\it after} the
transition.

The major difference with the weak bound is that, before the
transition, there is a single correlation length $\xi (t)$, since
the symmetry is unbroken. Further,as we shall see, the time ${\bar
t}$ after the transition occurs early, just before the order
parameter has reached its equilibrium value. At this time the real
and imaginary parts of $\phi$ are still effectively independent,
and we still have only one correlation length.

We reach the same conclusion from the viewpoint of the causal
horizon.  There is a time ${\bar t}$,
\begin{equation}
h({\bar t}) = \xi_{eq}({\bar t}), \label{causals}
\end{equation}
when the horizon is big enough to accommodate a defect. Before
this time, $0<t<{\bar t}$, the adiabatic approximation has
completely broken down, and the field is frozen in. Up to factors
$O(1)$, this matches the value of ${\bar t}$ given in
(\ref{causals}). It is argued that, at best, the system will
emerge at time ${\bar t}$ with a correlation length ${\bar\xi} =
\xi_{eq}({\bar t})$ that characterises identifiable domains.

More specifically, in this and the other approaches, it is the
longest wavelength modes, which control the field ordering, that
freeze in first and unfreeze last. It must be stressed that these
are rough estimates for the earliest time and the largest
correlation length when the system permits an adiabatic
description after the onset of the transition and we would not be
surprised if, in practice, there was a difference by a factor of a
few.

Whatever, in a second assumption, that is independent of
causality,
 Zurek further identified ${\bar\xi}={\xi_{eq}({\bar t})}$ with $\xi_{def} ({\bar t})$, the
defect (vortex) separation at the earliest time, ${\bar t}$, at
which they could be produced.
 To be
specific, and simple, we assume that
\begin{equation}
\frac{d\epsilon}{dt} = -\frac{1}{\tau_Q}
\label{tQ}
\end{equation}
in the vicinity of $t=0$, and adopt mean field critical indices
for the equilibrium correlation length $\xi_{eq} (t)$ and the
speed of sound ${\bar c}(t)$,
\begin{equation}
\xi_{eq} (t) = \xi_{0}\bigg|\frac{t}{\tau_{Q}}\bigg|^{-1/2},
\,\,{\bar c} (t) = c_{0}\bigg|\frac{t}{\tau_{Q}}\bigg|^{1/2},
 \label{crit}
\end{equation}
at this time, for suitable (zero-temperature) parameters $\xi_0$
and $c_0 = \xi_0/\tau_0$.

From (\ref{causals}) it follows that
\begin{equation}
{\bar t} \approx\sqrt{\tau_Q\tau_0}={\bar t}_Z  \label{bart}
\end{equation}
and
\begin{equation}
{\bar\xi} = \xi_{eq}({\bar t}_Z)\approx
\xi_0\bigg(\frac{\tau_Q}{\tau_0}\bigg)^{1/4}={\bar\xi}_Z.
\label{zcaus}
\end{equation}
In practice, even for the fastest transitions, we have $\tau_Q\gg
\tau_0$, whereby ${\bar\xi}_Z\gg\xi_0$. The variant of the causal
argument that has the field frozen in with this correlation length
at $t=-{\bar t}_Z$ suggests that $\xi (t)\approx {\bar\xi}_Z$ for
$-{\bar t}_Z\geq t\leq{\bar t}_Z$. In particular $\xi (0)\approx
{\bar\xi}_Z$.

From the second assumption we have that the {\it initial} vortex
density ${\bar n}_{def}$ =$n_{def}({\bar t})$ is
\begin{equation}
{\bar n}_{def} = O\bigg(\frac{1}{{\bar\xi}_Z^{2}}\bigg)=
\frac{1}{f^{2}\xi_{0}^{2}}\bigg(\frac{\tau_{0}}{\tau_{Q}}\bigg)^{1/2},
\label{ndef}
\end{equation}
where $f = O(1)$ estimates the fraction of defects per 'domain'.
Equivalently, the length of vortices in a box volume $v$ is
$O({\bar n}_{def}v)$.

\subsection{Relativistic Field Theory}

Let us assume global symmetry breaking again. We note that if
there had been no critical slowing down, as in relativistic QFT,
and ${\bar c}(t) = c$ is constant, then the 'sonic' horizon $h(t)$
of (\ref{sonic}) is replaced by the usual causal 'light' horizon, to give
\begin{equation}
\xi (t) \leq h(t)\approx 2ct
\end{equation}
for the phase correlation length. This constraint was used to
bound the production of monopoles\cite{kibble3} and cosmic
strings\cite{joao} (vortices) in the early universe.

On the other hand the strong causal bound (\ref{causals}) now
gives

\begin{equation}
{\bar t}_K ={\bar t} \approx (\tau_Q\tau^{2}_0)^{1/3}
\label{bartk}
\end{equation}
and
\begin{equation}
{\bar\xi}_K ={\bar\xi} = \xi_{eq}({\bar t}_K)\approx
\xi_0\bigg(\frac{\tau_Q}{\tau_0}\bigg)^{1/3}. \label{kcaus}
\end{equation}
In (\ref{kcaus}) is the common correlation length for early time,
leading to the estimate ${\bar n}_{def} = O(1/{\bar\xi}^2_K)$ for
the defect density at the time of their production.

 We stress that critical slowing down is not necessary for the
 existence of causal bounds. Its importance lies in the
 ameliorating effect it has on temperature inhomogeneities, whose
 effect is to diminish defect production\cite{kv}.
In the early universe temperature inhomogeneity is slight, unlike
in condensed matter physics experiments, where it is often
substantial, and critical slowing down is crucial.

This is all for flat space-time.

\subsection{Non-causal Mechanisms: The Ginzburg Regime}

In particular, what strikes us about the causal predictions is
that they are as universal as the mean field approximation is
valid. That, of course, is their main attraction. In the first
instance the basic distance and time scales $\xi_0$ and $\tau_0$
do not use any information about the magnitude of the order
parameter (i.e. the strength of the interactions). Even the
adoption of non-mean-field critical indices (necessary for $^4He$,
unnecessary for $^3He$) hardly changes the picture.

In stressing the primacy of causal horizons Zurek downplayed the
importance of thermal fluctuations, whose characteristic scale is
set by the Ginzburg temperature $T_G$ through $\epsilon (T_G)$,
independent of the quench rate and dependent on the microscopic
parameters of the theory. This is in contrast to the earlier
suggestion (by Kibble\cite{kibble1} and others\cite{joao}) that
thermal fluctuations in the Ginzburg regime might also lead to the
production of vortices, again at early times. The reason why the
Ginzburg regime might
 be important is the following: once we are below $T_{c}$,
the Ginzburg temperature $T_{G} <T_{c}$ signals the temperature
above which there is a significant probability for thermal
fluctuations between one degenerate groundstate and another on the
scale of the correlation length at that temperature. That is, the
thermal energy in such a fluctuation matches the energy required
to pass over the hump of the unstable minimum.

 Whereas,
above $T_{G}$ one might anticipate a population of 'domains',
fluctuating in and out of existence, at temperatures below $
T_{G}$ fluctuations from one minimum to the other become
increasingly unlikely. Thus field configurations with non-trivial
topology formed above $T_G$ could stabilise. Since the Ginzburg
regime does depend on the value of the order parameter, $\epsilon
(T_G)$ can vary hugely from system to system. It had been
suggested\cite{kibble1} that we identify
\begin{equation}
 \xi_{eq}(T_{G}) = \frac{\xi_0}{\sqrt{1-T_G/T_c}}
\end{equation}
 with
the scale $\bar\xi$ at which stable domains begin to form. This is
still invoked in cosmology\cite{joao}.

We shall see later why this is not the case. However, even if
thermal fluctuations are not the mechanism for the formation of
larger domains, they cannot be ignored completely. At the least
they are relevant to the formation of small 'domains', and to
wiggles in the boundaries of larger domains where defects are to
be found. As such, they can make the definition of defect density
scale invariant.

\section{ AN ALTERNATIVE APPROACH: INSTABILITIES AND (LINE) ZEROES}

In the several years since these simple bounds were first proposed
we have acquired a much better understanding of the way in which
transitions occur. These does not mean that these bounds have lost
their relevance, but that they need to be qualified.

To be more quantitative we assume that our {\it single} complex
scalar field $\phi$ in three spatial dimensions has a wine-bottle
potential. The transition is continuous. That is, we assume that
the qualitative dynamics are conditioned by the field's {\it
equilibrium} free energy, of the form
\begin{equation}
F(T) = \int d^{3}x\,\,\bigg(|\nabla\phi |^{2} + \epsilon(T)|\phi
|^{2} + \lambda |\phi |^{4}\bigg). \label{FR}
\end{equation}
where we have scaled the field to be as generic as possible.

On rescaling $\phi$ could represent a relativistic quantum field, with
free energy
\begin{equation}
F(T) = \int d^{3}x\,\,\bigg(|\nabla\phi |^{2} + m^2(T)|\phi
|^{2} + \lambda |\phi |^{4}\bigg). \label{QFT}
\end{equation}
where $m^2(T) = M^2\epsilon (T)$ at one-loop level.
On the other hand, on scaling, $F$ could be the Ginzburg-Landau
free energy
\begin{equation}
F(T) = \int d^{3}x\,\,\bigg(\frac{\hbar^{2}}{2m}|\nabla\phi |^{2}
+\alpha (T)|\phi |^{2} + \beta |\phi |^{4}\bigg) \label{FNR}
\end{equation}
in which the chemical potential $\alpha (T) = \alpha_{0}\epsilon
(T)$ vanishes at the critical temperature $T_{c}$.   With minor
qualifications (\ref{FNR}) is the free energy adopted for
simplified $^3 He$ in Ref.\onlinecite{kopnin} and Bose-Einstein
condensation\cite{anglin} (BEC). It is not a reliable
representation of $^4 He$, but may still be useful in this case.

\subsection{Time-scales: The Spinodal Time}

Any dynamical equations for the onset of a continuous transition
will embody causality, by definition. However, the transition
cannot be said to have happened before the order parameter has
achieved its equilibrium value $|\phi |^2 = 1/\lambda$, (or
$M^2/\lambda$ or $\alpha_0/\beta$, depending on how we rescale
(\ref{FR})). If $\langle ...\rangle_t$ denotes ensemble averaging
at time $t$ then a lower bound on the first time from which we can
start counting defects is $t=t_{sp}$, for which $\langle |\phi
|^2\rangle_t =1/\lambda$.

Ensemble averaging is taken with respect to the relative
probability $p_{t}[\phi]$ that the order parameter field $\phi$
takes the value $\phi ({\bf x})$ at time $t$. The way in which
$\langle |\phi |^2\rangle_t$ builds up to its final value is by
the growth of the amplitudes of the unstable long-wavelength
modes, which are unstable  because of the upturned parabolic free
energy at initial times. These long wavelength modes order the
field on increasingly larger scales. The time $t_{sp}$ is,
crudely, the time for these modes to roll from the top of the hill
to the groundstates at the bottom, and we have termed it the
spinodal time. This is a very different picture from that of a
field freezing in by, or before, the transition that was proposed
in Section 2.

A priori, $t_{sp}$ is not related to the causal ${\bar t}_K$ or ${\bar t}_Z$, but it
is not difficult to see why they might be comparable. Unstable
modes grow exponentially fast. As long as dimensional analysis
makes ${\bar t}_K$ or ${\bar t}_Z$ the natural unit in which to measure time, any
exponentially growing term will achieve values that are not
exponentially large at times $t = O(1)$ in these units. We thus
anticipate only a logarithmic sensitivity to the microscopic
parameters of the theory (and the quench rate).

\subsection{Quantum Field Theory: The Decoherence Time}

Although the spinodal time is equally important for QFT,
there are additional problems in that the evaluation of ensemble averages
 $\langle ...\rangle_t$ is not enough. In particular, in QFT we
need to consider the whole density matrix
$\langle\phi^+|{\hat\rho} (t)|\phi^- \rangle$ rather than just the
diagonal elements $p_{t}[\phi ] = \langle\phi |{\hat\rho} (t)|\phi
\rangle$ that determine the probability functional.

The reason is that probabilities are only useful when there is no,
or little, quantum interference between adjacent configurations
(cf. the two-slit problem). Such interference is measured by the
magnitude of the off-diagonal density matrix elements.

Before we can think of identifying particular field configurations
like vortices in QFT we must have that the reduced density matrix
is approximately diagonal on coarse-graining.

 By coarse-graining we now mean the separation of
the whole into the 'system', the long wavelength modes of $\phi$
which establish the field ordering, and its 'environment' whose
degrees of freedom are integrated over to produce the reduced
matrix. The environment comprises the other fields with which our
scalar is interacting, together with the short wavelength modes of
the scalar field itself \cite{muller2,lombardo}. This
coarse-graining is more sophisticated than the effective
short-distance cutoff that is essentially all that is needed in
condensed matter.

The reduced density matrix is non-unitary (dissipative) and the
effect of this environment is to introduce a new time scale, the
'decoherence' time $t_D$ such that quantum interference can be
ignored at times later than it. Only for $t>t_D$ does it make
sense to think of classical defects. However, dissipation will
also grow exponentially initially since it is driven by the same
long wavelength unstable modes. Thus, although $t_D$ depends on
interaction strengths as well as the quench rate, we anticipate
that it is insensitive to them, for the same reasons that ${\bar
t} $ is insensitive. For the moment we assume that $t_D\leq
t_{sp}$ to keep the discussion simple.

\subsection{Length-scales: Vortices as Line Zeroes}

As we noted earlier, this dynamic picture, in which modes build up
their amplitudes from their initial small Boltzmann values when
the field is sitting at the top of the hill, by rolling down it,
or spreading over it, seems totally at variance with that given in
Section 2. There, it was assumed that the field freezes in at, or
even prior to the transition, with a correlation length
${\bar\xi}$ (${\bar\xi}_K$ or ${\bar\xi}_Z$) that is already huge,
comparatively.

In fact, even at early times when we can get by with a single
correlation length for the $\phi$-field, we are talking about two
{\it different} lengths. The length ${\bar\xi}$ is the usual
correlation length, obtained from the long-distance behaviour of
the correlation function
 at time $t$,
\begin{equation}
\langle\phi ({\bf x})\phi^* ({\bf 0})\rangle_t = G (r,t) = \int d
\! \! \! / ^3 k \, e^{i {\bf k} . {\bf x} } P(k,t), \label{G^*}
\end{equation}
($r = |{\bf x}|$) in which the $P(k,t)$ is the power spectrum of
the fluctuations. To obtain ${\bar\xi}$ we take
$r\rightarrow\infty$, and look for behaviour of the form $\exp
-(r/{\bar\xi})^{\gamma}$. That is, ${\bar\xi}$ is determined by
the nearest singularity of $P(k,t)$ in the $k$-plane.

This is not the length that characterises the separation of
defects, which are formed from the phase separation as the field
falls from the hill in different directions at different spatial
points. Vortices form because these phases do not match up
smoothly and are frustrated by zeroes in the field.

More specifically, at long times after the transition, we expect
to find widely separated classical vortices. The classical
vortices of this simple theory, solutions to the equation $\delta
F/\delta\phi = 0$ are tubes of false groundstate, with line-zeroes
$\phi = 0$ at their cores, with width $O(\xi_{eq}(T))$. If we
write $\phi = (\phi_1 +\phi_2 )/\sqrt{2}$ these line-zeroes are
the intersections of the surfaces $\phi_1 =\phi_2 = 0$. From the
earliest numerical simulations onwards\cite{tanmay,andyjames}
simple vortices (and defects in lower dimension) have been
identified by counting zeroes of the fields.

That is, the empirical measure of defect separation is
$\xi_{zero}$, the typical separation of line-zeroes. If, for the
sake of argument, we assume Gaussian field fluctuations, then we
shall see that, at their time of formation, $\xi_{zero}(t)$ is
given by\cite{halperin,maz}
\begin{equation}
\xi^2_{zero}(t) = O\bigg(\frac{-G(0,t)}{G''(0,t)}\bigg).
\label{gaus}
\end{equation}
Primes denote differentiation with respect to $r$.
Empirically\cite{iba}, (\ref{gaus}) has been known to be valid
until the spinodal time, even though the field has ceased to be
Gaussian by then.

 This suggests that a more
realistic length-scale for calculating defect densities  at the
onset of defect production is ${\bar\xi}_{zero} =
\xi_{zero}(t_{sp})$. In this case there is no immediate reason to
relate ${\bar\xi}_{zero}$ to ${\bar\xi}$ since, unlike
$\xi_{eq}(t)$, $\xi_{zero}(t)$ is given entirely by the {\it
short}-distance behaviour of $G(r,t)$ or equivalently, by the
dominant wavenumbers in $P(k,t)$.

It is not difficult to extend the picture of vortices as line zeroes
to other defects like global monopoles and domain walls in three dimensions,
or kinks in one dimension. No novelty arises, and we shall not do so.

\subsection{When are Line Zeroes Vortices?}

We have rather assumed that, even if we cannot identify
${\bar\xi}_{zero}$ with $\xi_{eq}({\bar t})$ (${\bar t}_Z$ or ${\bar t}_K$), we can identify
${\bar\xi}_{zero}$ with ${\bar\xi}_{def}$.

This cannot be the case exactly in that, whereas all such defects
can be identified by their line-zeroes (or zeroes), not all line
zeroes (or zeroes) are candidate defects, since zeroes occur on
all scales. A starting-point for counting vortices is to count
line zeroes of an appropriately coarse-grained field, in which
structure on a scale smaller than $\xi_{0}$, the classical vortex
size, is not present\cite{popov}. Lattice-based numerical
simulations do this automatically (but see
Ref.\onlinecite{gleiser}). For the moment, we put in a cutoff $l =
O(\xi_{0})$ by hand. We note that the inclusion of a cut-off does
not affect the long-distance correlation ${\bar\xi}$.

Suppose that, at time $t_{sp}$, the typical separation of
line-zeroes for this coarse-graining is $\xi_{zero,l}(t_{sp})$.
Before line zeroes can be identified with classical (global)
vortex cores, and $\xi_{zero,l}(t_{sp})$ with $\xi_{def}(t_{sp})$,
the following conditions are necessary.

\begin{itemize}
\item
Although the correlation length $\xi (t)$ is insensitive to a
short-distance cutoff $l$, this is not the case for
$\xi_{zero,l}(t)$. Thermal fluctuations will give structure on
small scales which will lead to ambiguity in its definition and in
the subsequent density. Only when $\partial
\xi_{zero,\,l}/\partial l$ is small in comparison to
$\xi_{zero,\,l}/l$ at $l = \xi_{0}$ will the line-zeroes have the
small-scale non-fractal nature of classical defects, although
defects
 may behave like random walks on larger scales. Among other things,
 this will depend on the Ginzburg regime.
\item
As a final, related, check, the energy in field gradients should
be commensurate with the energy in classical vortices with the
same density as that of line zeroes.
\end{itemize}

In fact, most (but not all\cite{calzetta}) numerical lattice
simulations cannot distinguish between line-zeroes and classical
vortices.

We shall see that, roughly, fluctuations separate into thermal
fluctuations controlled by the current temperature $T(t)$, which
determine the small-scale structure of line-zeroes, which compete
with large-amplitude long-wavelength fluctuations whose role is to
provide large-scale order. It is these latter with which classical
defects are associated.

\subsection{Counting Zeroes}

Suppose, at some time, that the field has line zeroes ${\bf
R}_{n}(s)$, where $n = 1,2,..$ labels the zero, and $s$ measures
the length along it.  As a result the {\it topological line
density} of zeroes ${\vec{\rho}} (\bf r)$ can be
defined\cite{halperin,maz} by
\begin{equation}
\vec{\rho}({\bf x}) = \sum_{n}\int ds \frac{d{\bf R}_{n}}{ds}
\delta^{3} [{\bf x} - {\bf R}_{n}(s)]. \label{corrr}
\end{equation}
In (\ref{corrr}) $ds$ is the incremental length along the line of
zeroes ${\bf R}_{n}(s)$  and $d{\bf R}_{n}/ds$ is a unit vector
pointing in the direction which corresponds to positive winding
number.

It follows that, in terms of the zeroes of the field $\phi ({\bf
x})$, $\rho_{i}({\bf x})$ can be written as
\begin{equation}
\rho_{i}({\bf x}) = \delta^{2}[\phi ({\bf
x})]\epsilon_{ijk}\partial_{j} \phi_{1}({\bf x})
\partial_{k}\phi_{2}({\bf x}), \label{rho3}
\end{equation}
where $\delta^{2}[\phi ({\bf x})] = \delta[\phi_{1} ({\bf x})]
\delta[\phi_{2} ({\bf x})]$, where $\phi = (\phi_1 + i\phi_2)/\sqrt{2}$. The coefficient of the
$\delta$-function in (\ref{rho3}) is the Jacobian of the more
complicated transformation from line zeroes to field zeroes. What
we want is not this, but the {\it total line density}
$\bar{\rho}({\bf x})$,
\begin{equation}
\bar{\rho_{i}}({\bf x}) = \delta^{2}[\phi ({\bf
x})]|\epsilon_{ijk}\partial_{j} \phi_{1}({\bf x})
\partial_{k}\phi_{2}({\bf x})|. \label{rhobar}
\end{equation}

Let us suppose that, at time $t$, the probability density that the
field takes functional form $\phi({\bf x})$ is $p_t [\phi]$. The
ensemble average of $K[\phi]$ at time $t$ is
\begin{equation}
\langle K[\phi]\rangle_t = \int {\cal D}\phi_1{\cal D}\phi_2\,p_t
[\phi]\, K[\phi].
\end{equation}
The vanishing field expectation value and the  independence of the
field and its derivatives
\begin{equation}
\langle\phi_{a}({\bf x})\rangle_t = 0 = \langle\phi_{a}({\bf
x})\partial_{j}\phi_{b}({\bf x})\rangle_t , \label{gg4}
\end{equation}
imply $\langle\rho_{j}({\bf x})\rangle_t = 0$ {\it i.e.} an equal
likelihood of a string line-zero or an antistring line-zero
passing through an infinitesimal area. However,
\begin{equation}
n_{zero}(t) = \; \langle\bar{\rho_{i}}({\bf x})\rangle_{t} \; > 0
\label{n3}
\end{equation}
and measures the {\it total} line-zero density in the direction
$i$, without regard to string orientation.  The isotropy of the
initial state guarantees that $n$ is independent of the direction
$i$.

To get a basic idea as to what this means let us assume that the
field fluctuations are Gaussian. This has been the starting
assumption for networks of cosmic strings\cite{tanmay,andyjames}.
 Then everything is given in terms of the two-field correlator (\ref{G^*}) at
time $t$,
\begin{equation}
\langle\phi_{a} ({\bf x})\phi_{b} ({\bf 0})\rangle_t
=\delta_{ab}G({\bf r},t).\label{diag0}
\end{equation}
We coarse-grain the field by putting in a cutoff $l = O(\xi_{0})$
by hand, as
\begin{equation}
G_{l}(r, t) = \int d \! \! \! / ^3 k\, e^{i{\bf k}.{\bf x}}P(k,
t)\,e^{-k^{2}l^{2}}, \label{Gl}
\end{equation}
where we assume $l = O(\xi_0)$ or $O(\xi_{eq}(T_{final}))$, as
appropriate.

As we indicated earlier, in this Gaussian approximation
$n_{zero,l}(t)$ is determined completely\cite{halperin,maz} by the
{\it short-distance} behaviour of $G(r, t)$ as
\begin{equation}
n_{zero,l}(t) = \frac{1}{2\pi\xi_{zero,l}(t)^2}
=\frac{-1}{2\pi}\frac{G''_l(0, t)}{G_l(0, t )}, \label{ndeff}
\end{equation}
where we have used (\ref{ndeff}) to define $\xi_{zero}(t)$. That is,
$n_{zero}(t)$ is determined by the ratio of the fourth to second
moments of $P(k,t)$.

For non-Gaussian fields the situation is much more complicated.
However, as long as there is a dominant wavenumber $k_0(t)$ in
$P(k;t)$ this sets a length scale $\xi\approx k_0(t)^{-1}$ that
characterises vortex separation. As we said earlier, (\ref{ndeff})
can be approximately valid until the spinodal time, and we shall
assume that to be the case.

 We note that a line-zero is invariant under gauge
transformations $\phi\rightarrow\phi\,e^{i\alpha}$. Whereas a
simple spatial cutoff breaks gauge-invariance, we have learned
from lattice gauge theory how to coarse-grain a gauge field.
However, for a local theory, taking  $\phi_1$ and $\phi_2$
independently Gaussian is a gauge-dependent statement. We shall
not pursue this further.

We conclude with a brief discussion of topological charge through
a surface, for which the topological density (\ref{rho3}), rather
than the total density (\ref{rhobar}), is appropriate, again in
the Gaussian approximation. Consider a circular path in the bulk
material (in the 1-2 plane), circumference $C$, the boundary of a
surface $S$. For given field configurations $\phi_{a}({\bf x})$
the phase change $\theta_{C}$ along the path can be expressed as
the surface integral
\begin{equation}
\theta_{C} = 2\pi\int_{{\bf x}\in S} d^{2}x\,\,\rho_{3}({\bf x}),
\end{equation}
Again we quench from an initial state satisfying (\ref{gg4}) (i.e.
no rotation).

It is not difficult to show\cite{RKK,edik} that, if $f(r,t) =
G(r,t)/G(0,t)$, then
\begin{equation}
(\Delta\theta_{C})^{2}=\frac{C}{\xi_{s}(t)} =
2C\int_{0}^{\infty}dr\frac{f\prime_l^{2}(r,t)}{1 - f_l^{2}(r,t)}.
\label{step}
\end{equation}
The linear dependence on $C$ is purely a result of Gaussian
fluctuations. We note that $(\Delta\theta_{C})^{2}$ requires more
than the very short distance behaviour of $f_l$. Nonetheless, if
$P(k,t)$ is strongly peaked at $k = k_0$, then $\xi_s =
O(k_0^{-1})$, as is $\xi_{zero}$. Further, if we
removed all material except for a strip from the neighbourhood of the
contour $C$ we would still have the same result.  This supports
the assertion by Zurek that the correlation length for phase
variation in bulk fluid is also appropriate for annular flow.

That $(\Delta\theta_{C})^{2}$ varies in time is a consequence of
defects migrating through the boundary from the inside to the
outside and vice-versa, and annihilating. For a finite system,
where such migration is impossible and topological charge is
conserved, we would think of adopting (\ref{step}) at the moment
that defects appeared.

\section{WHEN DO THE APPROACHES AGREE? CONDENSED MATTER}

Let us summarise the situation for our idealised condensed matter,
as far as the 'strong' causal bounds of Zurek and Kibble are
concerned.. The time and length scales of Zurek's analysis are
${\bar t}_Z$ of (\ref{bart}) and ${\bar\xi}_Z=\xi_{eq}({\bar
t}_Z)$ of (\ref{zcaus}). In our analysis of zeroes the relevant
time and length scales are $t_{sp}$ and
${\bar\xi}_{zero}=\xi_{zero}(t_{sp})$. Above and beyond that,
thermal fluctuations may complicate the identity of
${\bar\xi}_{zero}$ and ${\bar\xi}_{def}$.

We have already suggested that ${\bar t}_Z$ and $t_{sp}$ may be comparable, qualitatively,
but the situation for length scales is less simple, not least because of thermal fluctuations.
To make further progress requires a concrete model in which correlation functions
can be calculated explicitly.

\subsection{The Time-Dependent Ginzburg-Landau (TDGL) Equation}

Not surprisingly, because the Time-Dependent Ginzburg-Landau
(TDGL) equation lends itself so directly to numerical analysis, it
has been the basis of many papers, largely by
Zurek\cite{zurek2,zurek3} and co-workers (but not exclusively e.g.
see Refs.\onlinecite{kopnin,calzetta}). By definition it encodes
causality. In Ref.\onlinecite{kopnin} a stripped-down global
$U(1)$ model was used to mimic the Helsinki $^3 He$
experiment\cite{helsinki}. A more realistic TDGL model for $^3
He$, that takes the full order-parameters into account has been
used elsewhere\cite{Bunkov}. The TDGL equation is also useful in
BEC\cite{anglin}.

In the last resort, only numerical methods permit us to get
long-time solutions, but in this article we pick out some
properties of simplified analytic solutions that we think show the
relevance  of the causal bounds.

We assume that, for the condensed matter systems of interest to
us, the dynamics of the transition can be derived from the
explicitly {\it time-dependent} Landau-Ginzburg free energy ($\xi_0 = 1$)
\begin{equation}
F(t) = \int d^{3}x\,\,\bigg(\frac{1}{2}(\nabla\phi_{a} )^{2}
+\frac{1}{2}\epsilon (t)\phi_{a}^{2} + \frac{1}{4}\lambda
(\phi_{a}^{2})^{2}\bigg). \label{F}
\end{equation}
in which we substitute $T(t)$ for $T$ directly in (\ref{FR}),
where $\phi = (\phi_{1} + i\phi_{2})/\sqrt{2}$ ($a=1,2$) is the
complex order-parameter field,

For a simple dissipative system the time-dependent Landau-Ginzburg
(TDLG) equation is
\begin{equation}
\frac{1}{\Gamma}\frac{\partial\phi_{a}}{\partial t} =
-\frac{\delta F}{\delta\phi_{a}} + \eta_{a}, \label{tdlg}
\end{equation}
where $\eta_{a}$ is Gaussian thermal noise, satisfying
\begin{equation}
\langle\eta_{a} ({\bf x},t)\eta_{b} ({\bf y},t')\rangle =
2\delta_{ab}T(t)\Gamma\delta ({\bf x}-{\bf y})\delta (t -t').
\label{noise}
\end{equation}

The basic timescale $\tau_0$, the relaxation time of the long
wavelength modes, is $\Gamma^{-1}$ in our units.

 In terms of the dimensionful constants of (\ref{FNR}) the fundamental length scale
$\xi_{0}$ is
 $\xi_{0}^{2} = \hbar^{2}/2m\alpha_{0}$ whereby
 $\xi_{eq}(T) = \xi_{0}(T/T_{c} -1)^{-1/2}$
as before.  With $\tau_{0} = 1 /\Gamma\alpha_{0}$ it follows that
the equilibrium correlation length $\xi_{eq} (t)$ and the speed of
sound behave when $t$ vanishes as in (\ref{crit}),
\begin{equation}
\xi_{eq} (t) = \xi_{0}\bigg|\frac{t}{\tau_{Q}}\bigg|^{-1/2},
\,\,{\bar c}(t) =
\frac{\xi_0}{\tau_0}\bigg|\frac{t}{\tau_{Q}}\bigg|^{1/2}.
\end{equation}
from which the causal bounds (\ref{bart}) and (\ref{zcaus})
follow.

\subsection{Freezing-in Happens (and is Irrelevant)}

It is relatively simple to show that the TDGL equation
(\ref{tdlg}) embodies Zurek's causality bound. At early times $
|t|\leq {\bar t}_Z$ the effective potential $V(\phi ,T)$ is still
roughly quadratic and  the self-interaction term can be neglected
($\lambda =0$).

In space, time
and temperature units in which $\xi_{0} = \tau_{0} = k_{B} =1$,
Eq.\ref{tdlg} then becomes
\begin{equation}
{\dot\phi}_{a}({\bf x},t) = - [-\nabla^{2} + \epsilon (t)]\phi_{a}
({\bf x},t) +{\bar\eta}_{a} ({\bf x},t). \label{free}
\end{equation}
where ${\bar\eta}$ is the renormalised noise. The solution of the
'free'-field linear equation is straightforward, giving a Gaussian
equal-time correlation function
\begin{equation}
G({\bf r},t) = \int d \! \! \! / ^3 k \, e^{i {\bf k} . {\bf r} }
P(k, t). \label{diag}
\end{equation}
in which the power spectrum $P(k,t)$ has a representation in terms
of the Schwinger proper-time $\tau$ as
\begin{equation}
P(k, t) = \int_{0}^{\infty} d\tau \,{\bar T}(t-\tau/2)\,e^{-\tau
k^{2}}\,e^{-\int_{0}^{\tau} ds\,\,\epsilon (t- s/2)},
\label{lgpower}
\end{equation}
where ${\bar T}$ is the renormalised temperature. In turn, this
gives\cite{ray2,ray3}
\begin{equation}
G(r, t) = \int_{0}^{\infty} d\tau\,{\bar T}(t-\tau/2)
\,\bigg(\frac{1}{4\pi\tau}\bigg)^{3/2}
e^{-r^{2}/4\tau}\,e^{-\int_{0}^{\tau} ds\,\,\epsilon (t- s/2)}.
\label{lgcorr}
\end{equation}

 For $\epsilon (t)$ of
Eq.\ref{tQ}  a saddle-point calculation shows that, at time $t=0$,
when the transition begins,
\begin{equation}
G(r,0)\approx \frac{T_c}{4\pi r}\,e^{-a(r/{\bar\xi}_Z)^{4/3}},
\label{notyuk}
\end{equation}
on rescaling, where $a = O(1)$, confirming Zurek's result of a
frozen field correlated over many cold defect thicknesses.

It is immediately apparent that ${\bar\xi}_Z)$ is not a measure of
the separation of line zeroes. Putting in the momentum cutoff
$k^{-1}> l =\bar{l}\xi_{0} =O(\xi_{0} )$ of Eq.\ref{lgcorr} by
hand corresponds to damping the singularity in $G(r,t)$ at $\tau =
0$ as\cite{ray2}
\begin{equation}
G_{l}(0,t)=\int_{0}^{\infty}\frac{ d\tau\,{\bar T}(t-\tau
/2)}{[4\pi (\tau + {\bar l}^{2})]^{3/2} }
e^{-r^{2}/4\tau}\,e^{-\int_{0}^{\tau} ds\,\,\epsilon (t-
s/2)},
\label{gfreel}
\end{equation}
making $G_{l}(0,t)$ finite. We stress that, for $t\approx 0$, the
correlation length ${\xi}$ remains $O({\bar\xi})$, {\it
independent} of $l$.

At $t=0$ both $G_{l}(0,t)$ and
\begin{equation}
-G''_l(0, t) = 2\pi\int_{0}^{\infty} d\tau\,
\,\frac{{\bar T}(t-\tau
/2)}{[4\pi (\tau + {\bar l}^2)]^{5/2}}\,e^{-\int_{0}^{\tau}
ds\,\,\epsilon (t- s/2)}. \label{lgcor3}
\end{equation}
are dominated by the short wavelength fluctuations at small
$\tau$. Even though the field is correlated over a distance
${\bar\xi}_Z\gg l$ the density of line zeroes $n_{zero} =
O(l^{-2})$ depends entirely on the scale at which we look.
Equivalently, $\xi_{zero}(0) = O(l)$. In no way would we wish to
identify these line zeroes with prototype vortices and
$\xi_{zero}(0)$ has nothing to do with ${\bar\xi}_Z$ at this time.

\subsection{Why $t_{sp}\approx {\bar t}_Z$ and
${\bar\xi}_{zero}\approx{\bar\xi}_{def}\approx {\bar\xi}_Z$ (in General)}

As the system evolves away from the transition time, the free
equation Eq.\ref{free} ceases to be strictly valid but, to a first
approximation, the back-reaction does not set in until the field
has sampled the groundstates. We can keep the linear
approximation, with its single correlation length, until $t\approx
t_{sp}$. For the linear quench of (\ref{tQ}) we find\cite{ray2}
\begin{equation}
\langle |\phi |^2\rangle_{t}= \,e^{2\int_{0}^{t}du\,|\epsilon (u)|
}\int_{0}^{\infty} d\tau\,{\bar T}(t-\tau/2)\frac{e^{-(\tau -
2t)^{2}|\epsilon '(t_{0})|/4}}{[4\pi (\tau + {\bar l}^{2})]^{3/2}
}.
\end{equation}
 However, as time passes the peak
of the exponential grows and $n_{zero}$ becomes increasingly
insensitive to $l$. How much time we have depends on the magnitude
of ${\lambda^{-1}}$, since once $G(0,t)$ has reached this value it
stops growing. Provided that the peak at $\tau = 2t$ dominates
over the thermal fluctuations at $\tau\approx 0$ then
\begin{equation}
\langle |\phi |^2\rangle_{t}\approx\,{\bar T}_c\,e^{(t/{\bar t}_Z)^{2}
}\int_{0}^{\infty} d\tau\,\frac{e^{-(\tau -
2t)^{2}/4{\bar t}_Z^2}}{[4\pi (\tau + {\bar l}^{2})]^{3/2}}.
\end{equation}
Since $G(0,t) = O(\exp((t/{\bar t}_Z)^{2})$ at early
times the backreaction is implemented extremely rapidly, justifying the
free-field approximation in the expression for
$G_{l}(0,t)$ above.

Let us assume that the thermal fluctuations can be ignored by time $t = {\bar t}_Z$.
In our units, in which $\lambda{\bar T}_c = \sqrt{1-T_G/T_c}$, the
condition that $\langle |\phi |^2\rangle_{t} = 1/\lambda$ at $t=t_{sp}$ gives
\begin{equation}
\frac{t_{sp}}{{\bar t}_Z}\approx\frac{1}{2}\sqrt{\ln \bigg(\frac{\tau_Q/\tau_0}{(1-T_G/T_c)^2}\bigg)}\,\,>\,1.
\label{tspt}
\end{equation}
Although greater than unity, the ratio is $O(1)$, extremely
insensitive to the parameters of the model, and to the quench
rate. For systems and experiments as widely different as the ones
for $^3 He$ and $^4 He$ that we shall discuss later, we have
$t_{sp}/{\bar t}_Z\approx 3$.

If we continue to assume that the thermal fluctuations at
$\tau\approx 0$ can be ignored, then we can use (\ref{gfreel}) and
(\ref{lgcor3}) to calculate the density of line zeroes. The result
is

\begin{equation}
{\bar n}_{zero}= n_{zero}(t_{sp})\approx \frac{{\bar t}}{8\pi
t_{sp} }\frac{1}{\xi_{0}^{2}}\sqrt{\frac{\tau_{0}}{\tau_{Q}}},
\label{primdef}
\end{equation}
in accord with (\ref{ndef}) of Zurek, where $f^2\approx 8\pi
t_{sp}/{\bar t}_Z = O(10^2)$. Equivalently, the line-zero
separation is
\begin{equation}
{\bar\xi}_{zero}\approx\xi_{zero}(t_{sp})= 2(t_{sp}/{\bar t}_Z)^{1/2}{\bar\xi} = O({\bar\xi}_Z),
\end{equation}
even less sensitive to the parameters of the model and the quench rate.

By assumption, ${\bar n}_{zero}$ is independent of the cutoff $l$
($l= O(1)$ in our units). The line-zeroes are on the verge of
being classical vortices since, once $\langle |\phi
|^2\rangle_{t}= 1/\lambda$, the Gaussian field energy, largely in
field gradients, is
\begin{equation}
{\bar F}\approx\langle\int_{V}
d^{3}x\,\frac{1}{2}(\nabla\phi_{a})^{2}\rangle= -VG''(0,t_{sp}),
\end{equation}
where $V$ is the spatial volume. This matches the energy
\begin{equation}
{\bar E}\approx V n_{def}(t)(2\pi G(0,t_{sp})) = -VG''(0,t_{sp})
\end{equation}
possessed by a network of classical global strings with density
${\bar n}_{zero}$, in the same approximation of cutting off their
logarithmic tails.

\subsection{Why Back-reaction may not Matter (Much)}

The full TDGL equation is
\begin{equation}
{\dot\phi}_{a}({\bf x},t) = - [-\nabla^{2} + \epsilon (t)+
\lambda|\phi ({\bf x},t)|^{2}]\phi_{a} ({\bf x},t) +{\bar\eta}_{a}
({\bf x},t). \label{full}
\end{equation}

In order to retain some analytic understanding of the way that the
density is such an ideal quantity to make predictions for, we
adopt the approximation of preserving Gaussian fluctuations by
linearising the self-interaction as
\begin{equation}
{\dot\phi}_{a}({\bf x},t) = - [-\nabla^{2} + \epsilon_{eff}
(t)]\phi_{a} ({\bf x},t) +{\bar\eta}_{a} ({\bf x},t),
\end{equation}
where $\epsilon_{eff} $ contains a (self-consistent) term
$O(\lambda\langle |\phi |^{2}\rangle)$. Additive
renormalisation is necessary, so that $\epsilon_{eff}\approx
\epsilon$, as given earlier, for $t\leq t_{0}$.

Self-consistent linearisation is the {\it only} usable
approximation in non-equilibrium QFT\cite{boyanovsky}, but is not
strictly necessary here, since numerical simulations that identify
line zeroes of the field can be made that use the full
self-interaction\cite{zurek2,zurek3}. However, to date none
address the questions we are posing here exactly, and until then
there is virtue in analytic approximations provided they are not
taken too seriously. Our discussion complements that of
Ref.\onlinecite{Boyanovsky} in which a self-consistently
linearised TDGL theory is also examined.

 The solution for $G_l(r,t)$ is a straightforward generalisation of
(\ref{lgcorr}),
\begin{equation}
G_{l}(r,t)=\int_{0}^{\infty}\frac{ d\tau\,{\bar T}(t-\tau
/2)}{[4\pi (\tau + {\bar l}^{2})]^{3/2} }
e^{-r^{2}/4\tau}\,e^{-\int_{0}^{\tau} ds\,\,\epsilon_{eff} (t-
s/2)},
\end{equation}
making $G_{l}(0,t)$ finite.

Assuming a {\it single} zero of $\epsilon_{eff} (t)$ at $t = 0$,
at $r=0$ the exponential in the integrand peaks at $\tau
={\bar\tau} = 2t$.  Expanding about ${\bar\tau}$ to quadratic
order gives
\begin{equation}
G_{l}(0,t)\approx \,e^{2\int_{0}^{t}du\,|\epsilon_{eff} (u)|
}\int_{0}^{\infty} d\tau\,{\bar T}(t-\tau/2)\frac{e^{-(\tau -
2t)^{2}|\epsilon '(t_{0})|/4}}{[4\pi (\tau + {\bar l}^{2})]^{3/2}
} .
\end{equation}
The effect of the back-reaction is to stop the growth of
$G_{l}(0,t)-G_{l}(0,0)= \langle |\phi |^{2}\rangle_{t}-\langle
|\phi |^{2}\rangle_{0}$ at its symmetry-broken value
$\lambda^{-1}$ in our dimensionless units, thereby preserving
Goldstone's theorem by requiring $|\epsilon_{eff}
(u)|\rightarrow\infty$.

What is remarkable in this approximation is that the density of
line zeroes uses {\it no} property of the self-mass contribution
to $\epsilon_{eff}(t)$, self-consistent or otherwise. With
\begin{eqnarray}
-G''_l(0, t) &=& 2\pi\int_{0}^{\infty} d\tau\,
\,\frac{{\bar T}(t-\tau
/2)}{[4\pi (\tau + {\bar l}^2)]^{5/2}}\,e^{-\int_{0}^{\tau}
ds\,\,\epsilon (t- s/2)}.
\\
&\approx& \,2\pi\,e^{2\int_{0}^{t}du\,|\epsilon_{eff} (u)|
}\int_{0}^{\infty} d\tau\,{\bar T}(t-\tau/2)\frac{e^{-(\tau -
2t)^{2}|\epsilon '(t_{0})|/4}}{[4\pi (\tau + {\bar l}^{2})]^{5/2}
}
\label{lgcorr3}
\end{eqnarray}
all prefactors in $n_{zero}$ cancel, to give\cite{ray3,ray}
\begin{equation}
n_{zero}(t) = \frac{1}{4\pi}\frac{\int_{0}^{\infty}
\frac{d\tau}{(\tau + {\bar l}^{2})^{5/2}} \,{\bar
T}(t-\tau/2)\,e^{-(\tau - 2(t-t_{0}))^{2}/4{\bar t}^{2}}}
{\int_{0}^{\infty} \frac{d\tau}{(\tau + {\bar l}^{2})^{3/2}}{\bar
T}(t-\tau/2)\,e^{-(\tau - 2(t- t_{0}))^{2}/4{\bar t}_Z^{2}}},
\end{equation}
thereby justifying our free-field approximation for $t<t_{sp}$. It
is for this reason that simple dimensional analysis (the basis of
the causal bounds) is so successful.

For $t>t_{sp}$ the equation for $n_{zero}(t)$ is not so simple
since the estimate above, based on a single isolated zero of
$\epsilon_{eff} (t)$, breaks down because of the approximate
vanishing of $\epsilon_{eff} (t)$ for $t>t_{sp}$.  A more careful
analysis shows that $G_{l}(0,t)$ can be written as
\begin{equation}
G_{l}(0,t)\approx \int_{0}^{\infty} \frac{ d\tau\,{\bar T}(t-\tau
/2)}{[4\pi (\tau + {\bar l}^{2})]^{3/2} } {\bar G}(\tau,t),
\end{equation}
where ${\bar G}(\tau,t)$ has the same peak as before at $\tau =
2t$, in the vicinity of which
\begin{equation}
{\bar G}(\tau,t) = e^{2\int_{0}^{t}du\,|\epsilon_{eff}
(u)|}\,e^{-(\tau - 2t)^{2}/4{\bar t}_Z^{2}},
\end{equation}
but ${\bar G}(\tau,t)\cong 1$ for $\tau < 2(t-t_{sp})$. Thus, for
$\tau_{Q}\gg\tau_{0}$, $G_{l}(0,t)$ can be approximately separated
as
\begin{equation}
G_{l}(0,t)\cong G_{l}^{UV}(t) + G^{IR}(t),
\end{equation}
where
\begin{eqnarray}
G_{l}^{UV}(t)&=& \int_{0}^{\infty} d\tau\,{\bar
T}(t-\tau/2)\,/[4\pi (\tau + {\bar l}^{2})]^{3/2} \label{GUV}
\\
&\approx& {\bar T}(t)\,\int_{0}^{\infty} d\tau\,/[4\pi
(\tau + {\bar l}^{2})]^{3/2}
\end{eqnarray}
describes  the
scale-{\it dependent} short wavelength thermal noise, proportional
to temperature, and
\begin{eqnarray}
G^{IR}(t) &=& \frac{1}{(8\pi t)^{3/2}} \,\int_{-\infty}^{\infty}d\tau
{\bar T}(t-\tau/2){\bar G}(\tau,t)
\\
&\approx &\frac{{\bar T}_{c}}{(4\pi{\bar\tau}(t))^{3/2}}
\,\int_{-\infty}^{\infty}d\tau {\bar G}(\tau,t)
\end{eqnarray}
describes the scale-{\it independent}, temperature independent,
long wavelength fluctuations. The integral $\int_{-\infty}^{\infty}d\tau {\bar
G}(\tau,t)$ is naturally time-dependent, largely cancelling its prefactor
so as to keep $\langle |\phi |^{2}\rangle_{t}$ constant.

A similar decomposition
$G\prime\prime_{l}(0,t)\cong G\prime\prime_{l}^{UV}(t) +
G\prime\prime^{IR}(t)$ can be performed as
\begin{eqnarray}
G\prime\prime_{l}^{UV}(t)&=& 2\pi\,\int_{0}^{\infty} d\tau\,{\bar
T}(t-\tau/2)/[4\pi (\tau + {\bar l}^{2})]^{5/2} \label{G''UV}
\\
&\approx& 2\pi{\bar
T}(t)\,\int_{0}^{\infty} d\tau\,/[4\pi (\tau + {\bar
l}^{2})]^{5/2}.
\end{eqnarray}
and
\begin{eqnarray}
G\prime\prime^{IR}(t) &=&\frac{4\pi}{(8\pi t)^{5/2}}
\,\int_{-\infty}^{\infty}d\tau {\bar T}(t-\tau/2)\,{\bar
G}(\tau,t)\\
&\approx&\frac{4\pi{\bar
T}_{c}}{(4\pi{\bar\tau}(t))^{5/2}} \,\int_{-\infty}^{\infty}d\tau
{\bar G}(\tau,t).
\end{eqnarray}
In particular, $G\prime\prime^{IR}(t)/G^{IR}(t)= O(t^{-1})$.

Firstly, suppose that, for $t\geq t_{sp}$,
  $ G^{IR}(t)\gg G_{l}^{UV}(t)$ and  $
G\prime\prime^{IR}(t)\gg G\prime\prime_{l}^{UV}(t)$, as would be
the case for a temperature quench ${\bar T}(t)\rightarrow 0$.
 Then, with little thermal noise, we have widely separated
line zeroes, with density $n_{zero}(t)\approx
-G\prime\prime^{IR}(t)/2\pi G^{IR}(t)$. With $\partial
n_{zero}/\partial l$ small in comparison to $n_{zero}/l$ at $l =
\xi_{0}$ we identify such essentially non-fractal line-zeroes with
prototype vortices, and $n_{zero}$ with $n_{def}$.  Of course, we
require non-Gaussianity to create true classical energy profiles.
Nonetheless, the Halperin-Mazenko result may be well approximated
for a while even when the fluctuations are no longer
Gaussian\cite{calzetta}. This is supported by the observation
that, once the line zeroes have straightened on small scales at
$t>t_{sp}$, the Gaussian field energy continues to match that
possessed by a network of classical global strings with density
$n_{zero}$.


\subsection{The Role of the Ginzburg Regime}

We have seen that the UV thermal fluctuations at time $t$ come
from the small $\tau$ part of the integration and are,
approximately, proportional to $T(t)$. It is the thermal
fluctuations that give rise to small-scale structure on the
defects and prevent us from giving a scale-independent value to
their density. Thus, if we reduce the temperature to absolute
zero, the thermal fluctuations vanish and the density becomes well
defined.

Our first observation is that, in practice, this never happens. In
pressure quenches it is the critical temperature that changes, at
almost constant temperature. Even in cooling, the final
temperature is usually a substantial fraction of the critical
temperature.

For the sake of simplicity, let us keep $T_f = O(T_c) < T_c$
 and thereby  take
$T=T_{c}$ in $G_{l}(0,t)$ above. The necessary time-{\it
independence} of $ G^{IR}(t)$ for $t>t_{sp}$ is achieved by taking
$\epsilon_{eff} (u)= O(u^{-1})$. In consequence, as $t$ increases
beyond $t_{sp}$ the relative magnitude of the UV and IR
contributions to $G_{l}(0,t)$ remains {\it approximately
constant}. Further, since for $t = t_{sp}$,
\begin{equation}
e^{2\int_{t_{0}}^{t}du\,|\epsilon_{eff} (u)|}\,e^{-(\Delta
t)^{2}/{\bar t}_Z^{2}}\approx 1,
\end{equation}
this ratio is the ratio at $t = t_{sp}$.

Nonetheless, as long as the thermal fluctuations are insignificant
at $t=t_{sp}$ the density of line zeroes will remain largely
independent of scale. This follows if $
G\prime\prime^{IR}(t_{sp})\gg G\prime\prime_{l}^{UV}(t_{sp})$,
since $G\prime\prime_{l}(0,t)$ becomes scale-independent later
than $G_{l}(0,t)$. In \cite{ray2} we showed that this is true
provided
\begin{equation}
(\tau_{Q}/\tau_{0})(1-T_{G}/T_{c})<C\pi^{4}, \label{ginlim}
\end{equation}
where $C= O(1)$ and $T_G$ is the Ginzburg temperature. That is,
for slow quenches or quenches in which thermal fluctuations remain
large, there are no identifiable defects at the spinodal time,
since the line zeroes are highly fractal on small scales and the
Zurek analysis breaks down.

In our self-consistent linearisation the situation never improves,
although this is extreme.
 Of course, if we view line-zeroes through
a lattice, they will be seen. Choose a different lattice, and we
will see more, or less. Numerical simulations that identify
vortices with line zeroes (and nothing more) on a fixed lattice
are suspect until thermal fluctuations become unimportant.
Certainly, for uniform quenches, defects are not formed at the end
of the Ginzburg regime by thermal fluctuations freezing in, as
originally suggested.

A simple way to see if defects can be produced by thermal
fluctuations is to keep the system in the Ginzburg regime for a
long period $\Delta t \gg t_{sp}$ and then drop out of it.
Numerical simulations have been performed by Zurek and
collaborators\cite{luis} that test this possibility. Empirically,
the end result is not very different from the case of the linear
quench that we have been discussing hitherto. The thermal
fluctuations of the Ginzburg regime are {\it not} the source of
defect production.

To get an analytic idea as to why this is the case we return to
the linear approximation of the previous section. Even with  more
complicated temperature profiles the separation into long and
short wavelength modes will still occur.

In particular, we expect the end-point behaviour  (\ref{GUV}) and
(\ref{G''UV}) for the scale-dependent short wavelength
fluctuations to be unchanged. As a result, $G_{l}^{UV}(t)$ and
$G\prime\prime_{l}^{UV}(t)$ are largely insensitive of the past
history of the system, even if that history involved a long time
$\Delta t$ in the Ginzburg regime.

Suppose, as in Ref.\onlinecite{luis}, we begin at temperature
$T_0$, then drop to temperature $T_G$  for a period $\Delta t$,
and then drop to temperature $T=0$. In comparison to spending the
whole time at $T_G$, the strength of the short wavelength
fluctuations in $G\prime\prime_{l}^{UV}(t)$ is
\begin{equation}
  O\bigg(\frac{1}{t^{3/2}}\bigg(\frac{T_0}{T_G}-\frac{\Delta
  t}{t}\bigg)\bigg)\approx 0
\end{equation}
for $t>>\Delta t$, with comparable behaviour for $G_{l}^{UV}(t)$.
That is, if we end up at low temperature, our intermediate history
is largely unimportant for the small-scale fluctuations, however
that temperature is arrived at.
 In particular, short-distance
scale-dependent fluctuations will be negligible at long times,
whatever.

 Similarly, provided the
temperature is monotonically decreasing (since otherwise the
effect on $G_{l}^{IR}(t)$ and $G\prime\prime_{l}^{IR}(t)$ is
dramatic) we do not expect a strong effect of past history on the
IR contributions beyond a delay in the position ${\bar\tau}$ of
the peak in $\tau$ by approximately $2\Delta\tau$.  Since
${\bar\tau}(t)$ depends on where $\epsilon_{eff}(t-\tau /2)$
vanishes, we must have
 ${\bar\tau}(t) = O(t)$, largely insensitive to intermediate conditions, once
 $t$ is large enough.
On neglecting the (now negligible) UV terms, we have $n_{zero}$
determined approximately by ${\bar\tau}(t)$ alone (giving the
usual $n_{zero}\propto t^{-1}$ behaviour). To a good approximation
this is what is seen in the numerical simulations\cite{luis}. To
quantify the slight differences seen in the numerical simulations
arising from different temperature profiles requires further work.

There is no way that defects can be produced by attempting to
incubate them in the Ginzburg regime.

\section{WHEN DO THE APPROACHES AGREE? RELATIVISTIC QFT}

For QFT the situation is rather different. In the previous
section, instead of working with the TDLG equation, we could have
worked with the equivalent Fokker-Planck equation\cite{luis2} for
the probability $p_{t}[\Phi ]$ that, at time $t>0$, the
measurement of $\phi$ will give the function $\Phi ({\bf x})$.
When solving the dynamical equations for a hot quantum field  it
is convenient to work with probabilities from the start.

\subsection{The Closed Time-path}

Take $t=t_0$ as our starting time for the evolution of the complex
field $\phi = (\phi_1 +i\phi_2)/\sqrt{2}$.  Suppose that, at this
time, the system is in a pure state, in which the measurement of
$\phi$ would give $\Phi_0({\bf x})$. That is:-
\begin{equation}
\hat{\phi}(t=0,{\bf x}) | \Phi_0,t=t_0 \rangle = \Phi_0 | \Phi_0,t=t_0
\rangle.
\end{equation}
The probability $p_{t}[\Phi]$ that, at time $t_f>t_0$, the
measurement of $\phi$ will give the value $\Phi$ is $p_{t}[\Phi] =
|\Psi_{0}|^2$, where $\Psi_{0}$ is the state-functional with the
specified initial condition.  As a path integral
\begin{equation}
\Psi_{0}  = \int_{\phi(t_0) = \Phi_0}^{\phi(t) = \Phi} {\cal D} \phi
\, \exp \biggl \{ i S_t [\phi] \biggr \},
\end{equation}
where $S_t [\phi]$ is the (time-dependent) action that describes
how the field $\phi$ is driven by the environment
 and spatial and field
labels have been suppressed (e.g. ${\cal D}\phi ={\cal
D}\phi_1{\cal D}\phi_2 )$. Specifically, for $t > t_0$ the action
for the field is taken to be
\begin{equation}
S_t [\phi] = \int dx \biggl ( \frac{1}{2} \partial_{\mu} \phi_{a}
\partial^{\mu} \phi_{a} - \frac{1}{2}\epsilon (t)M^2 \phi_{a}^2 -
\frac{1}{4} \lambda (\phi_{a}^2)^2 \biggr ). \label{St}
\end{equation}
where $\epsilon(t)$ is as before. Henceforth we return to natural
units in which $\xi_0$ = $\tau_0$ = $M^{-1}$ = $1$.

It follows that $p_{t}[\Phi]$ can be written in the closed
time-path form
\begin{equation}
p_{t}[\Phi] = \int_{\phi^{\pm}(t_0) = \Phi_0}^{\phi^{\pm}(t) =
\Phi} {\cal D} \phi^+  {\cal D} \phi^- \, \exp \biggl \{ i \biggl
( S_t [\phi^+]-S_t [\phi^-] \biggr ) \biggr \},
\end{equation}
where ${\cal D}\phi^{\pm} ={\cal D}\phi^{\pm}_1{\cal
D}\phi^{\pm}_2 $. Instead of separately integrating $\phi^{\pm}$
along the time paths $t_0 \leq t \leq t_f$, the integral can be
interpreted as time-ordering of a field $\phi$ along the closed
path $C_+ \oplus C_-$ where $\phi =\phi^+$ on $C_+$ and $\phi=
\phi^-$ on $C_-$.
 When we extend
the contour from $t_f$ to $t= \infty$ either $\phi^+$ or $\phi^-$
is an equally good candidate for the physical field, but we choose
$\phi^+$.

The choice of a pure state at time $t=t_0$ is too simple to be of
any use. Let us assume that $\Phi$ is Boltzmann distributed at
time $t=t_0$ at an effective temperature of $T_0 = \beta_0^{-1}$
according to the  Hamiltonian $H[\Phi]$ corresponding to the
action $S[\phi ]$, in which $\phi$ is taken to be periodic in
imaginary time with period $\beta_0$. We now have the explicit
form for $p_{t}[\Phi]$,
\begin{equation}
p_{t} [ \Phi] = \int_B {\cal D} \phi \, e^{i S_C [\phi]} \, \delta
[ \phi^+ (t_f) - \Phi ],
\end{equation}
 written as the time ordering of a
single field along the contour $C=C_+ \oplus C_- \oplus C_3$,
extended to include a third imaginary leg, where $\phi$ takes the
values $\phi^+$, $\phi^-$ and $\phi_3$ on $C_+$, $C_-$ and $C_3$
respectively, for which $S_C$ is $S[\phi^+]$, $S[\phi^-]$ and
$S_0[\phi_3]$.
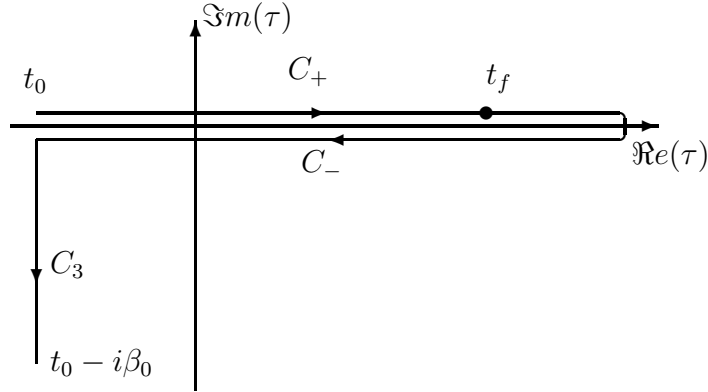
\begin{figure}[htb]
\begin{center}
\setlength{\unitlength}{0.5pt}
\begin{picture}(495,280)(35,480)
\put( 70,565){\makebox(0,0)[lb]{\large $C_3$}}
\put(260,640){\makebox(0,0)[lb]{\large $C_-$}}
\put(185,750){\makebox(0,0)[lb]{\large $\Im m(\tau)$}}
\put(250,710){\makebox(0,0)[lb]{\large $C_+$}}
\put(510,645){\makebox(0,0)[lb]{\large $\Re e (\tau)$}}
\put(400,705){\makebox(0,0)[lb]{\large $t_f$}} \put(
50,705){\makebox(0,0)[lb]{\large $t_0$}}
\put(70,490){\makebox(0,0)[lb]{\large $t_0-i \beta_0$}}
\thicklines \put(400,690){\circle*{10}} \put( 40,680){\vector( 1,
0){490}} \put( 60,690){\vector( 1, 0){220}} \put(280,690){\line(
1, 0){220}} \put(500,680){\oval(10,20)[r]}
\put(500,670){\vector(-1, 0){220}} \put(280,670){\line(-1,
0){220}} \put( 60,670){\vector( 0,-1){110}} \put( 60,560){\line(
0,-1){ 60}} \put(180,480){\vector( 0, 1){280}}
\end{picture}
\end{center}
\caption{The closed timepath contour $C_+ \oplus C_-$ with a third
imaginary leg}
\end{figure}
It is sufficient for our purposes to take the Boltzmann
distribution on $C_3$ to be the {\it free-field} distribution with
mass parameter $m^2 = M^2\epsilon_0$, corresponding to a
temperature quench from an initial state at temperature $T_{0}>T_c
$, where $(T_{0}/T_c -1) = \epsilon_{0}$.

As with condensed matter, it is not necessary to calculate
$p_t[\Phi]$ directly since $G_{ab}(|{\bf x} -{\bf x} '|;t) =
\langle\Phi_a ({\bf x})\Phi_b ({\bf x}')\rangle_{t}$ is given by
\begin{equation}
G_{ab}(|{\bf x} -{\bf x} '|;t) = \langle\phi_a ({\bf x},t)\phi_b
({\bf x}',t)\rangle, \label{wight}
\end{equation}
the equal-time thermal Wightman function with the given thermal
boundary conditions.

\subsection{The (Irrelevant) Freeze-in}

Unlike for condensed matter we shall not postulate dissipative
equations (with corresponding noise), but just calculate the
evolution of the quantum field under its time-dependent
Hamiltonian derived from (\ref{F}). Insofar as the results look
similar it is for totally different reasons.[For simplicity we
continue to work in flat space-time.]

As for the condensed matter case, the interval $-{\bar t}_K \leq
t\leq {\bar t}_K$ occurs in the {\it linear} regime, when the
self-interactions are unimportant. The relevant equation for
constructing the correlation functions of this one-field system is
now the second-order equation
\begin{equation}
\frac{\partial^{2}\phi_a}{\partial t^{2}} = -\frac{\delta
F}{\delta\phi_a}, \label{op}
\end{equation}
for $F$ of Eq.\ref{FR}. This is solvable in terms of the mode
functions $\chi^{\pm}_{k}(t)$, identical for $a=1,2$, satisfying
\begin{equation}
\Biggl [ \frac{d^2}{dt^2} + {\bf k}^2 + \epsilon (t) \Biggr
]\chi^{\pm}_{k}(t)  =0, \label{mode}
\end{equation}
subject to $\chi^{\pm}_{k}(t)
=
e^{\pm i\omega_{in}t}$ at large negative times, with incident
frequency $\omega_{in} = \sqrt{{\bf k}^{2} +\epsilon_{0}}$.

The diagonal correlation function $G(r, t)$ of Eq.\ref{diag} is
given as the equal-time propagator
\begin{equation}
G(r, t)=\int d \! \! \! / ^3 k \, e^{i {\bf k} . {\bf x} }
\chi^{+}_{k}(t) \chi^{-}_{k}(t)C(k)
\label{modesum}
\end{equation}
where $ C(k) =\coth(\omega_{in} (k)/2T_0 )/2\omega_{in}(k)$
encodes the initial conditions.

An exact solution can be given\cite{bowick} in terms of Airy
functions. Dimensional analysis shows that, on ignoring the
k-dependence of $C(k)$, appropriate for large $r$ (or small $k$),
${\bar\xi}_K=\xi_{eq}({\bar t}_K)$ of Eq.\ref{kcaus} again sets
the scale of the equal-time correlation function. Specifically,
\begin{equation}
G(r,0)\propto\int d\kappa\,\frac{\sin\kappa
(r/{\bar\xi}_K)}{\kappa (r/{\bar\xi}_K)}\,F(\kappa ), \label{GQFT}
\end{equation}
where $F(0) = 1$ and $F(\kappa )\sim \kappa^{-3}$ for large
$\kappa$.

Kibble's prediction for the large scale at which the field freezes
in is correct but, as with condensed matter, has nothing to do
with the formation of defects around $t=0$. At $t=0$ there are, as
yet, no unstable modes and, as far as defects are concerned the
field becomes ordered, as before, because of the exponential
growth of long-wavelength modes, which stop growing once the field
has sampled the groundstates.  What matters is the relative weight
of these modes in $P(k,t)$ (the 'Bragg' peak) to the fluctuating
short wavelength modes, since the contribution of these latter is
very sensitive to the cutoff $l$ at which we look for defects.
Only if their contribution to Eq.\ref{ndef} is small when field
growth stops can a network of vortices be well-defined at early
times. At $t=0$ there is no Bragg peak and the density of line
zeroes depends entirely on the scale at which we are looking.

\subsection{The Spinodal Time and $\xi_{zero}$}

We begin by extending the analysis from $t=0$ to later times,
still in the approximation of a {\it free} roll. This needs care
for slow quenches since the back-reaction serves to hold the field
in the vicinity of the intermediate ground-states $|\phi^{2}| =
\phi_{0}^{2}(t)$ where, now
\begin{equation}
\phi_{0}^{2}(t) = \frac{-\epsilon (t)}{\lambda} =
\frac{1}{\lambda}\frac{ t}{\tau_Q}
\end{equation}
for $t<\tau_Q$.

Prior to the completion of the quench at $ t = \tau_Q$, the
mode equation (\ref{mode}), now of the form
\begin{equation}
\Biggl [ \frac{d^2}{dt^2} + {\bf k}^2  -\frac{t}{\tau_Q} \Biggr ]\chi^{\pm}_{k}(t)  =0,
\label{mode1s}
\end{equation}
is exactly solvable, as we saw earlier.
 As in the
case of condensed matter previously, we have coarse-grained the
field by introducing a simple cut-off at $k = \Lambda = O(M)$ in
dimensional units, or $l = \Lambda^{-1}$. In natural units ${\bar
l}= O(1)$.  The unstable exponentially growing modes appear when
\begin{equation}
\Omega_{k}^{2}(t) = -{\bf k}^2  +\frac{t}{\tau_Q}>0.
\end{equation}
For fixed $k$ this occurs when $t>t_{k} = \tau_Qk^{2}$ (or $k^2 =
k_t^2= t/\tau_Q $).  As before, we are not looking for vortices
within vortices, but coarse-grain to the cold vortex size.

The WKB solution is adequate for our purposes. $G_{l}(r;t)$ can be
written as $G_{l}(r;t) = G^{exp}(r;t)+G_{l}^{osc}(r;t)$ where
\begin{equation}
G^{exp}(r;t)\simeq T_0\int_{|{\bf k}|<k_t} d \! \!
\! / ^3 k\frac{S_{k}(t)}{|k_{t}^{2}-k^{2}|^{1/2}} \,e^{i {\bf k}
. {\bf x}}|\alpha^{+}_{k}I_{1/3}(S_{k}(t))+\alpha^{-}_{k}
I_{-1/3}(S_{k}(t))|^{2} \label{WI}
\end{equation}
has exponentially growing long wavelength modes and
\begin{equation}
G_{l}^{osc}(r;t)\simeq T_0\int_{\Lambda >|{\bf
k}|>k_t} d \! \! \! / ^3
k\frac{S_{k}(t)}{|k_{t}^{2}-k^{2}|^{1/2}} \,e^{i {\bf k} . {\bf
x}
}|\alpha^{+}_{k}J_{1/3}(S_{k}(t))-\alpha^{-}_{k}J_{-1/3}(S_{k}(t))|^{2}
\label{WJ}
\end{equation}
has short wavelength oscillatory modes. In both cases
\begin{equation}
S_{k}(t) = \int^{t}_{t^{-}_k}dt'\,|\Omega_{k}(t')|
=\frac{2}{3}\frac{1}{\sqrt{\tau_Q}}|t-t_k |^{3/2}.
 \label{Sk}
\end{equation}
We stress that, at the level of a free field (or any
self-consistent  Gaussian) inserting a cut-off is identical to
{\it integrating} out the short wavelength modes, for an unbiased
initial state. Provided we are far from the transition we have
incorporated the initial data into the $\alpha^{\pm}$ in
(\ref{WI}) and (\ref{WJ}), which have no $\lambda$ or $\tau_Q$
dependence.

For large $t$ the integrand in (\ref{WI}) will have its 'Bragg'
peak at some $k_{0}(t)\rightarrow 0$ as $t\rightarrow \infty$,
once the angular integrals have been performed. Assuming that
$k_{0}(t)\ll k_t$ the upper bound in the integral can be dropped
and $|k_{t}^{2}-k^{2}|$ approximated by $|k_{t}^{2}|$, knowing
that there is no singularity at $k=k_t$. With nothing to stop
$|\alpha^{+}_{k}+ \alpha^{-}_{k}|^{2}$ behaving like a nonzero
constant in the vicinity of $k=0$, it can be treated as slowly
varying and the integral approximated as\cite{KMR}
\begin{eqnarray}
G^{exp}(r;t)&\propto &
T\bigg(\frac{\tau_Q}{t}\bigg)^{1/2}
e^{(4/3)(t/{\bar t}_K)^{3/2}}\int_{|{\bf k}|<1} d \! \! \! / ^3 k
\, e^{i {\bf k} . {\bf x} } \;e^{-2\sqrt{t\tau_Q}k^{2}}
\nonumber
\\
&\propto &
\frac{T}{|\epsilon(t)|^{1/2}}\bigg(\frac{1}{\sqrt{t\tau_Q}}\bigg)^{3/2}e^{(4/3)(t/{\bar
t}_K)^{3/2}} e^{-r^{2}/2\xi^{2}(t)} \label{Wexp}
\end{eqnarray}
on performing the $k^{2}$ expansion of the exponent, where
\begin{equation}
\xi^{2}(t) = \sqrt{t\tau_Q} \label{chis}
\end{equation}
in natural units $\tau_0 = 1$. Although, like Eq.\ref{Wexp}, the
expression Eq.\ref{chis} is not supposed to be valid for small
$t$, it does embody the Kibble freezeout condition  in satisfying
${\dot\xi}({\bar t}) = O(1)$.

There are two immediate differences between (\ref{Wexp}) and its
condensed matter counterpart (\ref{lgcorr}). Firstly, in
(\ref{Wexp}) the length $\xi (t)$ of (\ref{chis}) is {\it both}
the correlation length of the field {\it and} exactly the
separation of line zeroes $\xi (t)_{zero}$ as defined through
(\ref{ndeff}).  Secondly, insofar that the time-dependence of $\xi
(t)$ is a guide to early evolution, it grows as $t^{1/4}$, rather
than the more customary $t^{1/2}$ of condensed matter.

The calculations above were for a free roll. Let us suppose,
provisionally, that the backreaction exerts its influence over
such a short time that, in effect, it is if it were an {\it
instantaneous} brake to domain growth. The provisional freeze-in
time $t^{*}$ is then, effectively, the time it takes to reach the
transient groundstate $\phi_{0}^{2}(t)$. That is, $G(0;t^{*}) =
O(\phi_{0}^{2}(t^{*}))$, giving
\begin{equation}
(\sqrt{t^{*}\tau_Q})^{3/2}e^{-(4/3)(t^{*}/{\bar t}_K)^{3/2}} =
O\bigg(\lambda^{1/2}(\frac{\tau_Q}{t^{*}})^{3/2}\bigg).
\label{tsff}
\end{equation}
This gives
\begin{equation}
\frac{t^*}{{\bar t}_K}\approx
\frac{1}{2}\bigg[\ln\bigg(\frac{\tau_Q/\tau_0}{(1-T_G/T_c)^2}\bigg)\bigg]^{2/3},
 \label{tfs}
\end{equation}
and we have used $1-T_G/T_c = \lambda$, to be specific. For small
coupling $t^*/{\bar t}_K >1$, but will be $O(1)$. A comparison
with (\ref{tspt}) shows strong similarities with condensed matter
and that, yet again, $t^*$ is insensitive to the parameters of the
theory.
 As far as the separation of scales is
concerned, we have the same effect qualitatively if we had taken
$t^{*}$ as the time for the field to reach the final ground state
as $|\phi^{2}| = 1/\lambda$, rather than the provisional
ground states $\phi_{0}^{2}(t)$.  Hence we can identify $t^*$ with $t_{sp}$ to
a good approximation and we shall not distinguish between them.

At this qualitative level the correlation length, and line-zero
separation length at the spinodal time is
\begin{equation}
\xi^{2}(t_{sp})= \xi^{2}(t_{sp})_{zero}\simeq {\bar\xi}^2_K
(\frac{t_{sp}}{{\bar t}_K})^{1/2} = O({\bar\xi}^2_K).
\label{chiss2}
\end{equation}

\subsection{Fluctuations}

All of this assumes that the oscillatory wavelength terms
(\ref{WJ}) can be ignored. Although we have adopted a cutoff at
$l=O(\xi_0) = O(M^{-1})$, there is still a contribution from modes
near $M^{-1}$. When we take these modes into account the density
of line zeroes at $t_{sp}$ can be written as
\begin{equation}
{\bar n}_{zero}= \frac{1}{f^2}{\bar n}_{K} (1 + E). \label{nisMf}
\end{equation}

The error term $E =O(\lambda^{1/2}(1/\tau_Q)^{4/3}(\ln
(1/\lambda))^{-1/3})$ is due to oscillatory modes, sensitive to
the cut-off $\Lambda$. The condition $E^{2}\ll 1$, necessary for
the line-zero density to be insensitive to scale, is
satisfied\cite{ray} if
\begin{equation}
(\tau_{Q}/\tau_{0})^{2}(1-T_{G}/T_c)<C, \label{ginlim2}
\end{equation}
where $C = O(1)$. This is the QFT counterpart to Eq.\ref{ginlim}.
If this is the case then $f^2 = 2\pi \sqrt{t_{sp}/{\bar t}_K}$,
that shows yet again that the causal estimate can be correct, but
for different reasons.

\subsection{Backreaction in QFT}

To improve upon the free-roll result we adopt a mean-field
approximation along the lines of \cite{boyanovsky,LA}, as we did
for the condensed matter systems earlier.

$G(r;t)$ still has the mode decomposition of (\ref{modesum}), but
the modes $\chi^{\pm}_{k}$ now satisfy the equation
\begin{equation}
\Biggl [ \frac{d^2}{dt^2} + {\bf k}^2 + \epsilon(t) +
\lambda\langle\Phi^{2}({\bf 0})\rangle_{t} \Biggr
]\chi^{\pm}_{k}(t) =0, \label{modeh}
\end{equation}
where we have taken $N=2$ in a large-N $(O(N))$ theory.

The end result is\cite{boyanovsky} that, on making a single
subtraction at $t=0$,
\begin{equation}
\Biggl [ \frac{d^2}{dt^2} + {\bf k}^2 + \epsilon (t) + \lambda\int
d \! \! \! / ^3 p \, C(p) [\chi^{+}_{p}(t)\chi^{-}_{p}(t)-1]
\Biggr ]\chi^{\pm}_{k}(t) =0, \label{modeh2}
\end{equation}
which we write as
\begin{equation}
\Biggl [ \frac{d^2}{dt^2} + {\bf k}^2 -\epsilon_{eff}(t) \Biggr
]\chi_{k}(t)  =0. \label{modemu}
\end{equation}
On keeping just the unstable modes in $\langle\Phi^{2}({\bf
0})\rangle_{t}$ then, as it grows, its contribution to
(\ref{modeh2}) weakens the instabilities, so that only longer
wavelengths become unstable. At $t^{*}$ the instabilities shut
off, by definition, and oscillatory behaviour ensues.

Explicit calculation shows that the backreaction has little effect
for times $t<t_{sp}$. For $t>t_{sp}$ oscillatory modes take over
the correlation function and we expect oscillations in $G(k;t)$.

In practice the backreaction rapidly forces $\epsilon_{eff}(t)$
towards zero if the coupling is not too small\cite{boyanovsky}. In
that case the end result is a new power spectrum, obtained by
superimposing oscillatory behaviour onto the old spectrum. As a
gross oversimplification, the contribution from the earlier
exponential modes alone can only be to contribute terms something
like
\begin{eqnarray}
G(r;t)&\propto&
\frac{T}{|\epsilon(t_{sp})|^{1/2}}e^{(4/3)(t_{sp}/{\bar t}_K)^{3/2}}
\int_{|{\bf k}|<1} d \! \! \! / ^3 k \, e^{i {\bf k} . {\bf x} }
\;e^{-2\sqrt{t_{sp}\tau_Q}k^{2}} \nonumber
\\
&\times&\bigg[\cos k(t-t_{sp}) + A(k)\sin k(t-t_{sp})\bigg]^{2}
\label{Wexps}
\end{eqnarray}
to $G$, where the details of $A(k)$ do not concern us.

The $k=0$ mode of Eq.\ref{Wexps} encodes the simple solution
$\chi_{k=0}(t) = a + bt$ when $\mu^2 = 0$. This has weak causality
built into it.
Specifically, for $r,t\rightarrow\infty$,
\begin{equation}
G(r,t)\approx \frac{C}{r}\theta (2t/r -1).
\end{equation}
It has to be said that this approximation should not be taken very
seriously for large $t$, since we would expect rescattering to
take place at times $\Delta t = O(1/\lambda)$ in a way that the
Gaussian approximation precludes.

However, it demonstrates that weak causality, implemented by the
Goldstone particles of the self-consistent theory, is likely to
have little effect on the density of line-zeroes that we expect to
mature into fully classical vortices, depending as it does largely
on the behaviour of $G$ at $r=0$.

Finally, all of the above was predicated on the long wavelengths
of the field having decohered by $t_{sp}$.  It is to this we now
turn.

\section{QFT: THE DECOHERENCE TIME}

The empirical TDGL condensed matter Langevin equation (\ref{tdlg})
that we adopted earlier could have been rewritten as a
Fokker-Planck equation\cite{luis2} for the functional probability
density $p_t[\phi]$. Our discussion of QFT in the previous Section
also encoded the corresponding $p_t[\phi]$. The difference is
that, for QFT, we have yet to ascertain whether we can use
$p_t[\phi]$ to count classical configurations, like defects. The
reason was given earlier: classical defects cannot be identified
until adjacent path histories decohere (i.e. until quantum
interference is negligible), since only then will $p_t[\phi]$
(approximately) obey classical laws.

The mechanism for decoherence is the interaction of the field with
its environment. In practice, all fields to which the $\phi$-field
couple will help it decohere. In addition, the short wavelength
modes, which play no role in field ordering and subsequent defect
production, are an important part of the environment. For
exemplary purposes we consider only the effects of short
wavelengths of the $\phi$-field itself on
decoherence\cite{muller2,lombardo}. A more complete discussion
that includes the effects of additional fields has been given
elsewhere\cite{lmr}. At the level that we have handled QFT so far
(a self-consistent Gaussian) our cut-off of short wavelength modes
is the same as integrating them out, and there is no decoherence.
We shall need to go beyond a free-field approximation to see any
effects.

Decoherence is measured in terms of dissipation. A byproduct of
establishing decoherence is that we recover a Fokker-Planck-like
equation for $p_t[\phi]$ or, equivalently, a generalised Langevin
stochastic equation for the classical field. Together with this
should be included the explicit dissipation in the early universe
as a consequence of the expansion of space-time. We shall not
treat that here, despite its importance (see
Refs.\onlinecite{ray,zurek4,stephens}).

To show the main ideas we shall consider the simpler system of a
single real field (whose defects are domain walls) undergoing a
very rapid quench. If we take the quench time
$\tau_Q\approx\tau_0$ this is equivalent to taking an
instantaneous quench, and this we do. There is no difficulty in
taking slower quenches in principle, it is just that the
calculations have yet to be performed.

That is, yet again we divide the field $\phi (x)$ into the long
wavelength 'system'-field $\phi_< (x)$ and the short wavelength
'environment'-field $\phi_> (x)$,
\begin{equation}
\phi_< (x) = \int_{|{\bf k}|< 1} d^3x\,\phi ({\bf x},t)\,e^{i{\bf k}.{\bf x}},\,\,\,\,
\phi_> (x) = \int_{|{\bf k}|> 1} d^3x\,\phi ({\bf x},t)\,e^{i{\bf k}.{\bf x}}.
\end{equation}
$|{\bf k}|= 1$ is the dividing line between stable and unstable
modes. With $\tau_Q\approx\tau_0$ (\ref{ginlim2}) is automatically
satisfied and varying the separation wavelength by a small amount
will have no effect.

Since it is the system-field $\phi_<$ whose behaviour changes
dramatically on taking $T$ through $T_{\rm c}$, we adopt an {\it
instantaneous} quench  for $T$ from $T_0$ to $T_{\rm f}=0$ at time
$t=0$, in which $\epsilon (T)$ changes sign and magnitude instantly,
concluding with the value $\epsilon (t)=-1, t>0$. Meanwhile, for
simplicity the $\phi_>$ mass is held at the original value $\epsilon_0$.

The classical action separates as
\begin{equation}
S[\phi ] = S_{\rm syst}[\phi_< ] + S_{\rm env}[\phi_> ] +
S_{\rm int}[\phi_< ,\phi_> ],
\end{equation}
where
\begin{eqnarray}
&&S_{\rm syst}[\phi_< ] = \int d^4x\left\{ {1\over{2}}\partial_{\mu}
\phi_<\partial^{\mu} \phi_< - {1\over{2}} \epsilon (t) \phi_<^2 -
{\lambda\over{4!}}\phi_<^4\right\}, \nonumber \\
&&S_{\rm env}[\phi_>
] = \int d^4x\left\{ {1\over{2}}\partial_{\mu}\phi_>
\partial^{\mu}
\phi_> - {1\over{2}}\epsilon_0 \phi_>^2 -
{\lambda\over{4!}}\phi_>^4\right\}, \nonumber
\\
&&S_{\rm int}[\phi_< ,\phi_> ] = - \frac{\lambda}{4} \int d^4x
\phi_<^2 (x) \phi_>^2 (x).
\nonumber
\end{eqnarray}
We have not included $\phi_<^3\phi_>$ and $\phi_<\phi_>^3$ terms
in $S_{\rm int}[\phi_< ,\phi_> ]$ since the former does not
contribute to very long wavelengths and the latter only appears at
higher order in $\lambda$. Our answers will have little to do with
the more familiar calculations of decoherence, which are motivated
by quantum mechanical systems with {\it linear} coupling to an
environment.

Beginning from an initial thermal distribution peaked around $\phi
= 0$ we follow the evolution of the system under the influence of
the short wavelength environment.  The spinodal time is now
 $ t_{\rm sp} \sim \frac{1}{2}\ln [1
 /\lambda ]$.

\subsection{The Reduced Density Matrix}

Whereas the field probabilities are just the diagonal elements of
the density matrix ${\hat\rho}$, decoherence is determined from
the full density matrix
\begin{equation}
\rho[\phi_<^+,\phi^+_> ,\phi^-_<,\phi^-_>,t]= \langle\phi^+_<
\phi^+_>\vert {\hat\rho} \vert \phi^-_< \phi^-_>\rangle.
\end{equation}
Our Gaussian initial conditions give an uncorrelated thermal
initial behaviour ${\hat\rho}[T_0] = {\hat\rho}_{<}[T_0]
{\hat\rho}_{>}[T_0]$ at temperature $T_0$.

The relevant object is the reduced density matrix. It describes
the evolution of the system under the influence of the
environment, and is defined by
\begin{equation}
\rho_{{\rm r}}[\phi^+_<,\phi^-_<,t] = \int {\cal D}\phi_{>}
~ \rho[\phi^+_<,\phi_{>} ,\phi^-_<,\phi_{>} ,t].
\end{equation}
The environment will have had the effect of making the system
essentially classical once $\rho_{r}(t)$ is, effectively,
diagonal.

Its temporal evolution is given by
\begin{equation}
\rho_{\rm r}[\phi_{{\rm
<f}}^+,\phi_{{\rm <f}}^-,t]= \int
d\phi_{{\rm <i}}^+ \int d\phi_{{\rm <i}}^- J_{\rm r}[\phi_{{\rm
<f}}^+,\phi_{{\rm <f}}^-,t\vert \phi_{{\rm <i}}^+,\phi_{{\rm
<i}}^-,t_0] ~~\rho_{\rm r}[\phi_{{\rm <i}}^+ \phi_{{\rm <i}}^-,t_0],
\end{equation}
where $J_{\rm r}$ is the reduced evolution operator
\begin{equation}
J_{\rm r}[\phi_{{\rm <f}}^+,\phi_{{\rm <f}}^-,t\vert \phi_{{\rm
<i}}^+,\phi_{{\rm <i}}^-,t_0] = \int_{\phi_{{\rm
<i}}^+}^{\phi_{{\rm <f}}^+} {\cal D}\phi^+_< \int_{\phi_{{\rm
<i}}^-}^{\phi_{{\rm <f}}^-} {\cal D}\phi^-_< ~e^{i\{S[\phi^+_<] -
S[\phi^-_<]\}} ~ F[\phi^+_<,\phi^-_<]. \label{evolred}
\end{equation}
The Feynman-Vernon\cite{feynver} influence functional
$F[\phi^+,\phi^-]$ is defined as
\begin{eqnarray}
& & F[\phi^+_<,\phi^-_<] = \int d\phi^+_{{\rm >i}} \int
d\phi^-_{{\rm > i}} ~ \rho_{\phi}[\phi_{{\rm > i}}^+,\phi_{{\rm >
i}}^-,t_0] \int d\phi_{{\rm > f}}\nonumber \\ & \times &
\int_{\phi^+_{{\rm > i}}}^{\phi_{{\rm > f}}}{\cal D}\phi^+_{\rm >}
\int_{\phi^-_{{\rm  > i}}}^{\phi_{{\rm > f}}} {\cal D}\phi^-_{\rm
>} \exp{\left(i\{S[\phi^+_{\rm >} ] + S_{{\rm int}}
[\phi^+_<,\phi^+_{\rm >} ]\right)}\nonumber\\ & \times &
\exp{\left(- i\{S[\phi^-_{\rm >}] + S_{{\rm
int}}[\phi^-_<,\phi^-_{\rm >}]\} \right)}, \nonumber
\end{eqnarray}
where we have used the closed time-path of Fig.1. From the
influence functional we define the influence action $\delta
A[\phi^+,\phi^-]$ as $F[\phi^+_<,\phi^-_<] = \exp {i \delta
A[\phi^+_<,\phi^-_<]}$.

We calculate to one loop. After defining $$\Delta
={1\over{2}}(\phi^{+2}_< - \phi^{-2}_<) ~~~,~~~ \Sigma
={1\over{2}}(\phi^{+2}_< + \phi^{-2}_<);$$ and using simple and
well known identities between propagators, the real and imaginary
parts of the influence action can be written as
\begin{equation}
{\rm Re} \delta A = \frac{\lambda^2}{4} \int d^4 x\int d^4y \Delta (x) K_{\rm
q}(x-y) \Sigma (y),\label{realpartIA}
\end{equation}
\begin{equation}
{\rm Im} \delta A = - \frac{\lambda^2}{8} \int d^4x\int d^4y \Delta (x) N_{\rm
q} (x,y) \Delta (y),\label{imaginarypartIA}
\end{equation}
where $K_{\rm q} (x-y) = {\rm Im} G_{++}^2(x,y) \theta (y^0-x^0)$
is the dissipation kernel and $ N_{\rm q} (x-y) = {\rm Re}
G_{++}^2(x,y)$ is the noise (diffusion) kernel.

\subsection{The Master Equation}

The first step in the evaluation of the evolution equation for
${\hat\rho}$, {\it the master equation}, is the calculation of the
density matrix propagator $J_{\rm r}$ from Eq. (\ref{evolred}). We
first perform a saddle point approximation
\begin{equation}
J_{\rm r}[\phi^+_{\rm <f},\phi^-_{\rm <f},t\vert\phi^+_{\rm
<i},\phi^-_{\rm <i}, t_0] \approx \exp{ i A[\phi^+_{\rm
cl},\phi^-_{\rm cl}]},
\end{equation}
where $\phi^\pm_{\rm cl}$ is the solution of the equation of
motion ${\delta Re
A\over\delta\phi^+_<}\vert_{\phi^+_<=\phi^-_<}=0$ with boundary
conditions $\phi^\pm_{\rm cl}(t_0)=\phi^\pm_{\rm <i}$ and
$\phi^\pm_{\rm cl}(t)=\phi^\pm_{\rm <f}$. Next, we assume that the
system-field contains only one Fourier mode with $\vec k = \vec
k_0$, which we identify with the Bragg peak when it arises.

To estimate $t_D$ it is sufficient to calculate the correction to
the usual unitary evolution coming from the noise kernel. For
clarity we drop the suffix $f$ on the final state fields. If
$\Delta = (\phi^{+2}_< - \phi^{-2}_<)/2$ for the {\it final} field
configurations, then the master equation for $\rho_r(t) = \langle
\phi^+_<\vert{\hat \rho}\vert\phi^-_<\rangle$ is

\begin{equation}
i {\dot \rho}_{\rm r} = \langle \phi^+_<\vert [{\hat
H},{\hat\rho_{\rm r}}] \vert \phi^-_<\rangle -
i\frac{\lambda^2}{8} V \Delta^2 D(k_0, t) \rho_{\rm r}+ ...
\label{master}
\end{equation}
The volume factor $V$  arises because we are considering a density
matrix which is a functional of two different field
configurations, $\phi^\pm_<(\vec x) = \phi^\pm_< \cos \vec k_0 .
\vec x$ spread over all space. For $k_0 < 1$ the diffusion
coefficient $D(k_0,t)$  shows the inevitable exponential growth
($T_0 = O(T_c)$)
\begin{equation}
D(k_0, t) \sim T_c ^2 \Omega (k_0)\exp [2\Omega (k)t],
\label{unstD2}
\end{equation}
where $\Omega^2(k) = M^2 - k^2$, (or $1-k^2$ in natural units),
associated with the instability of the $k_0$ mode. Eq.\ref{unstD2}
is only valid once $2\Omega (k)t>1$, which will have happened by
$t=t_{sp}$. Once $t>t_{\rm sp}$ the diffusion coefficient stop
growing, and oscillates around $D(k_0,t=t_{\rm sp})$.

On the other hand, for short times $ t \ll 1$ we find that
\begin{equation}
D(k_0, t) \sim T_c^2 t\label{unstD}
\end{equation}
up to coefficients ${\cal O}(1)$.

\subsection{The Decoherence Time}

We estimate $t_{\rm D}$ for the model by considering the
approximate solution to the master equation (\ref{master}),
\begin{equation}
 \rho_{\rm r}[\phi^+_<, \phi^-_<; t] \approx
\rho^{\rm u}_{\rm r}[\phi^+_<, \phi^-_<; t] ~ \exp
\bigg[-V\Gamma\int_0^t ds ~D(k_0, s) \bigg],
\end{equation}
where $\rho^{\rm u}_{\rm r}$ is the solution of the unitary part
of the master equation (i.e. without environment). In terms of the
fields $\bar\phi = (\phi_<^+ + \phi_<^-)/2,$ and
$ \delta = (\phi_<^+ - \phi_<^-)/2$, $\Gamma = (\lambda^2/8)
\bar\phi^2\delta^2$.

The system will decohere when the non-diagonal elements of the
reduced density matrix are much smaller than the diagonal ones.
We therefore look at the ratio
\begin{eqnarray}
\left\vert \frac {\rho_{\rm r}[\bar\phi+\delta,\bar\phi-\delta;t]}
{\rho_{\rm r}[\bar\phi,\bar\phi;t]}
\right\vert & \approx & \left\vert \frac {\rho_{\rm r}^{\rm u}[\bar\phi+\delta,
\bar\phi-\delta;t]}
{\rho_{\rm r}^{\rm u}[\bar\phi,\bar\phi;t]}
\right\vert\nonumber\\
&\times& \exp
\bigg[-V\Gamma\int_0^t ds ~D(k_0, s) \bigg ]\, .
\label{ratio}\end{eqnarray}

In general, it is not possible to obtain an analytic expression
for the ratio of density matrices that appears in Eq.(\ref{ratio})
since we do not even know the diagonal matrix well. However, since
diagonalisation will be found to occur for $t<t_{sp}$ it is
sufficient to neglect the self-coupling of the system field in the
diagonal matrix elements. In this case the unitary density matrix
remains Gaussian at all times as

\begin{equation}\left\vert \frac {\rho_{\rm r}^{\rm u}[\bar\phi+\delta,
\bar\phi-\delta;t]} {\rho_{\rm r}^{\rm u}[\bar\phi,\bar\phi;t]}
\right\vert = \exp [-T_{\rm c}\delta^2 p^{-1}(t)] , \label{uratio}
\end{equation}
where $p^{-1}(t)$, essentially\cite{guthpi,kim,gill} $
\langle\phi^2\rangle_t^{-1}$, decreases exponentially with time to
a value ${\cal O}(\lambda)$. In the unitary part of the reduced
density matrix the non-diagonal terms are not suppressed. [This
should not be confused with the observation that the unitary
Gaussian density matrix does show classical correlation, whereby
the Wigner functional becomes localised in phase space about its
classical solutions\cite{kim}. However, this has nothing to do
with eliminating quantum interference between different field
histories.] In order to obtain classical behaviour, the relevant
part of the reduced density matrix is the term proportional to the
diffusion coefficient in Eq.(\ref{ratio}), since it is this that
enforces its diagonalisation.

The decoherence time  $t_D$  can be defined as the solution to
\begin{equation}
1\approx V\Gamma \int_{0}^{t_{D}} ds ~D(k_0,
s).
\end{equation}
It corresponds to the time after which we are able to distinguish
between two different field amplitudes (inside a given volume
$V$).

To estimate $t_D$ we have to fix the values of $V$, $\delta$, and
$\bar\phi$. $V$ is understood as the minimal volume inside which
we do not accept coherent superpositions of macroscopically
distinguishable states for the field. Thus, our choice is that
this volume factor is ${\cal O}(M^{-3})= {\cal O}(1)$ in
dimensionless units,  since the Compton wavelength is the smallest
scale at which we need to look. Inside this volume, we do not
discriminate between field amplitudes which differ by $ {\cal
O}(1) $, and therefore we take $\delta \sim {\cal O}(1)$. For
$\bar\phi$ we set $\bar\phi^2={\cal O}(1 /\lambda)$, its
post-transition value.

From the short-times expression for the diffusion coeffient Eq.(\ref{unstD})
is very easy to show that decoherence does not occur while
$\mu t \ll 1$ due to the small value of the coupling constants.
Consequently, in order to evaluate the decoherence
time in our model, we have to use
Eq.(\ref{unstD2}). We obtain
\begin{equation}\exp [2 t_D] = O(1)
\end{equation}
from which it follows  that
 \begin{equation}
 1 < t_D < t_{\rm sp}.
\label{tdtsp}
\end{equation}

This result is welcome in that it suggests that our previous
analysis can probably go ahead to slower quenches. Moreover, if
(\ref{tdtsp}) is valid there we anticipate that classical defects
can appear as soon as the spinodal time is reached. However, some
caution is required since perturbation theory is, generally, not
valid for non-equilibrium processes. Fortunately, within the same
one-loop framework it is not difficult to find circumstances in
which (\ref{tdtsp}) holds until $t_{sp}$. In particular if, in
addition to the short wavelength modes, we include a large number
of weakly coupled external fields to decohere the long wavelength
modes (without any linear couplings), the result (\ref{tdtsp})
survives. Details are given elsewhere\cite{lmr}.

Finally, once the long wavelength modes have decohered they satisy
stochastic equations of a generalised Langevin
form\cite{muller2,lombardo,gleiser2}
\begin{equation}
\partial^2 \phi (x) - \phi + {\lambda\over{6}}\phi^3(x) +
{\lambda^2\over{2}} \phi (x)\int d^4y\,K(x,y)\phi^2(y)=
\lambda\phi (x)\eta (x), \label{lange}
\end{equation}
where the Gaussian noise $\eta$ satisfies $\langle\eta (x)\eta
(0)\rangle\propto N(x)$ in this approximation.

Since $K$ is retarded the integral describes Landau damping.
Because of the instantaneous nature of the quench we shall not
pursue this equation beyond saying that, even given that, with its
weak $O(\lambda)$ multiplicative noise and weak quasi-dissipative
term, it is very different from the generalised relativistic
Langevin equation invoked in Refs.\onlinecite{zurek2,zurek3}.

\section{EXPERIMENTS: CONDENSED MATTER}

Now that we understand the Kibble-Zurek bounds better, we give a brief
review of the main experimental evidence, as well as suggesting further
experiments that could provide further support for them.

Our first observation is that, in QFT, there is no evidence for or
against Kibble's bounds. Although we have every confidence in the
standard model of Electroweak unification (and its breaking) in
the early universe, we only have very circumstantial evidence for
phase transitions at the higher energies that are of more interest
to us for defect formation. In particular, to date there are no
reliable sightings of any topological defect that could have been
produced in the early universe. At best we have possible indirect
evidence, such as the suggestion that very high energy cosmic rays
arise from the intersection and release of energy from the high
energy density cores of cosmic strings. [There is a parallel in
condensed matter in quasiparticle production due to the
interaction of vortices in $^4 He$]. The original hope that
large-scale structure was determined by cosmic strings (vortices)
seems unfounded\cite{joao}.

On the other hand, for condensed matter systems there are several experiments that
have been devised to check Zurek's predictions.

\subsection{Superfluid $^3 He$}

We begin with superfluid $^3 He$ since, to date, it provides the
strongest support for the validity of Zurek's bounds. $^{3}He$ is
a {\it fermion} but, somewhat as in a BCS superconductor, these
fermions form the counterpart to Cooper pairs.  However, whereas
the (electron) Cooper pairs in a superconductor form a $^{1}S$
state, the $^{3}He$ pairs form a $^{3}P$ state. The order
parameter $A_{\alpha i}$ is a complex $3\times 3$ matrix
$A_{\alpha i}$. There are two distinct superfluid phases,
depending on how the global $SO(3)\times SO(3)\times U(1)$
symmetry is broken. If the normal fluid is cooled at low
pressures, it makes a transition to the $^{3}He-B$ phase, in which
$A_{\alpha i}$ takes the form $A_{\alpha i} = R_{\alpha
i}(\omega)e^{i\Phi}$, where $R$ is a real rotation matrix,
corresponding to a rotation through an arbitrary
$\omega$\cite{volovik}. Although more complicated than other
systems it can be easier to count vortices, since one can use NMR
to detect them.

So far, experiments have been of two types.  In the Helsinki
experiment\cite{helsinki} superfluid $^{3}He -B$ in a rotating
cryostat is bombarded by slow neutrons.  Each neutron entering the
chamber releases 760 keV, via the reaction $n + ^{3}He\rightarrow
p + ^{3}He + 760 keV$.  The energy goes into the kinetic energy of
the proton and triton, and is dissipated by ionisation, heating a
region of the sample above its transition temperature.  The heated
region then cools back through the transition temperature,
creating vortices. Vortices above a critical size  grow and
migrate to the centre of the apparatus, where they are counted by
an NMR absorption measurement.  The quench is very fast, with
$\tau_{Q}/\tau_{0} = O(10^{3})$.   Agreement with Eq.\ref{ndef} is
good.

 The second type of experiment has been performed at Grenoble
and Lancaster\cite{grenoble}.  Rather than count individual
vortices, the experiment detects the total energy going into
vortex formation. As before, $^{3}He$ is irradiated by neutrons.
After each absorption the energy released in the form of
quasiparticles is measured, and found to be less than the total
760 keV. This missing energy is assumed to have been expended on
vortex production.  Again, agreement with Zurek's prediction
Eq.\ref{ndef}  is good.

\subsection{$^4 He$ Experiments}

In $^{4}He$ the bose superfluid has as its order parameter a
complex field $\phi$, whose squared modulus $|\phi |^{2}$ is the
superfluid density. The global $U(1)$ symmetry breaks to yield
simple vortices, whose winding number measures the superflow
around them.

The experiments in $^{4}He$, conducted at Lancaster, follow
Zurek's original suggestion.  A sample of normal fluid helium, in
a container with bellows, is expanded so that it becomes
superfluid at essentially constant temperature. That is, we change
$1-T/T_c$ from negative to positive by increasing $T_c$. As the
system goes into the superfluid phase a network of vortices is
formed, which are detected by measuring the attenuation of second
sound within the bellows.  A mechanical quench is slow, with
$\tau_{Q}$ some tens of milliseconds, and $\tau_{Q}/\tau_{0} =
O(10^{10})$. Two experiments have been performed. In the
first\cite{lancaster} fair agreement was found with the prediction
Eq.\ref{ndef}. However, there were potential problems with
hydrodynamic effects at the bellows, and at the capillary with
which the bellows were filled.  A second
experiment\cite{lancaster2}, designed to minimise these and other
problems has failed to see any vortices whatsoever.

However, unlike the experiments with $^3 He$, the experimental
prediction for $^4 He$ needs more than causality to go further.
The density of vortices $n$  is
assumed to obey Vinen's equation\cite{vinen}
\begin{equation}
\frac{\partial n_{def}}{\partial t} = -\chi_{2}\frac{\hbar}{m}
n_{def}^{2}.
\end{equation}
 This behaviour, requiring $\xi_{def}(t)\propto t^{1/2}$ is the
 general behaviour for non-conserved order
parameters. We stress that causality does not provide a value for
$\chi_2$.

In our simple TDGL model, in which this behaviour follows at early
times in the absence of thermal fluctuations, $\chi_{2} =
4\pi\hbar\Gamma = 4\pi\hbar /\alpha_{0}\tau_{0}$. Taking
$\tau_{0}\approx 8.0\times 10^{-12}s$ and $\xi_{0}\approx 5.6\AA$
the mean-field approximation for $^{4}He$ gives $\chi_{2}\approx
5\times 4\pi$. [It would have the value $4\pi$ if we motivated
$\tau_{0}$ from the Gross-Pitaevskii equation\cite{zurek1}, in
which $\alpha_{0}\tau_{0} =\hbar$.] This decay law is assumed in
the analysis of the Lancaster experiments, in which the empirical
value of $\chi_2$  was not taken from quenches, but turbulent flow
experiments. It was suggested\cite{lancaster2} that $\chi_{2}
\approx 0.005$, orders of magnitude smaller than our suggestion
above. Although the TDLG theory is not very reliable for $^{4}He$,
if our estimate is at all sensible it does imply that vortices
produced in a {\it temperature} quench decay much faster than
those produced in turbulence.

  As a separate observation, we note
that the large value of $f^2$ in the prefactor of $n_{zero}$ is,
in itself, almost enough to make it impossible to see vortices in
$^{4}He$ experiments, should they be present.

To compound the problem, for early time at least, thermal
fluctuations are large in the Lancaster experiments. With
$\tau_{Q}/\tau_{0} = O(10^{10})$ and a Ginzburg regime  so large
that $(1-T_G /T_c ) = O(1)$ the inequality (\ref{ginlim}) is
hugely violated. In such circumstances the density of zeroes
$n_{zero}= O(l^{-2})$ after $t_{sp}$ depends exactly on the scale
$l$ at which we look and they are not candidates for vortices.
Since the whole of the quench takes place within the Ginzburg
regime this is not implausible. However, the Ginzburg-Landau
theory is not reliable for $^4He$ and, even though the thermal
noise never switches off, there is probably no more than a
postponement of vortex production.

\subsection{Superconductors: Flux Through a Surface}

The situation with {\it local} symmetries, in which the
$\phi$-field interacts with the electromagnetic fields, is
different, and only beginning to be understood\cite{mark}. Whereas
the 'domains' of the global symmetry breaking are characterised by
approximately constant field phases,  for a local (gauge) symmetry
the phase can be changed at will by a gauge transformation (except
at a zero of the field where it is not defined). Phase correlation
lengths are not physical observables. Nonetheless, vortices are
present, identified by their magnetic flux.

For the relativistic case discussed in Ref.\onlinecite{mark} it is
Landau damping that controls the relaxation time, commensurate
with our earlier comments on stochastic equations. Details are
given elsewhere in these proceedings\cite{arttu}. In this case it
is not clear if causality provides a useful constraint, except in
the case of extremely weak electromagnetic coupling, very fast
transitions or low temperatures\cite{zurek2} when, for condensed
matter, we have (\ref{zcaus}) for the density of Abrikosov
vortices\cite{calzetta}.

Even if the situation were clear here, there is a separate issue
as to how we could confirm any predictions.  Since total flux
(proportional to the number of vortices {\it minus} the number of
antivortices) is the gauge-invariant measurable quantity, rather
than the density (the number of vortices {\it plus} the number of
antivortices), we have to infer the separation ${\bar\xi}_{def}$
differently.

We assume a temperature quench of a superconductor with zero
average flux. Consider a closed path in the superconductor with
circumference $C\gg {\bar\xi}$. The phase difference
$\theta_C$ around the path is gauge invariant and we estimate
 the r.m.s phase difference along the path as
\begin{equation}
\Delta\theta_{C} = O(\sqrt{C/{\bar\xi}_{def}}).
\label{rphase}
\end{equation}
The variance of magnetic flux within the path is then
\begin{equation}
\Delta\Phi_{C}  =
O\bigg(\frac{\Phi_0}{2\pi}\sqrt{\frac{C}{{\bar\xi}_{def}}}\bigg),
\label{rflux}
\end{equation}
where $\Phi_0 = h/2e$ is the basic unit of magnetic flux. Since
topological charge is conserved, there is no haste to measure the
trapped flux. Such an experiment has been performed\cite{technion}
with a high-$T_c$ superconductor, but there is no agreement with
(\ref{rflux}) from its simple estimates. The reason is unclear.

In this context we should also mention the production of vortices
produced in the wake of Bose-Einstein condensation, for which the
Zurek mechanism also applies\cite{anglin}, but for which the
relevant experiments have yet to be performed.

\subsection{Annular experiments}

As an alternative way to check the predictions for ${\bar\xi}_{def}$ by
measuring conserved topological charge, Zurek suggested
experiments with annuli of condensed matter. If we quenched a
superconducting ring of circumference $C$ we would expect the
variance in flux through it to be given also by (\ref{rflux}),
provided its width is is less than ${\bar\xi}_{def}$, and assuming that
the fluctuations in the electromagnetic field are negligible.

In fact, it is more convenient to quench annular Josephson
Junctions, in which two rings of  superconductor are held apart by
an oxide layer through which Cooper pairs can tunnel. Let the
complex scalar Higgs fields for the superconductors (labeled $1$
and $2$) be $\phi_1$ and $\phi_2$ respectively. Despite our
concerns about the local symmetries within the individual
superconductors enabling us to gauge transform away the field
phases $\theta_1$ and $\theta_2$, the difference between these
phases determines a physical observable, the Josephson current,
\begin{equation}
J = J_c \sin\theta ,
\end{equation}
where $\theta = \theta_1 -\theta_2$.

 The defects of this system are the 'fluxons' of the
Josephson Junction and are easy to observe
experimentally\cite{monaco}. The variance in fluxon number at
their formation is expected to be
\begin{equation}
\Delta N_{C}=
\frac{1}{2\pi}\Delta\theta\approx\frac{1}{2\pi}\sqrt{\frac{C}{{\bar\xi}_{Z}}},
\label{fluxon}
\end{equation}
where ${\bar\xi}_{Z}$ is Zurek's causal length for such a system.
There are two competing mechanisms. At very early times we can
treat the two superconducting annuli as independent. If this
decoupling lasts long enough this would suggest\cite{RKK} that
$(\Delta\theta )^2=(\Delta\theta_1 )^2+(\Delta\theta_2 )^2$, for
which we use the Zurek causal constraints for free propagators. On
the other hand, if the coupling between the JTJs is important by
the causal time ${\bar t}_Z$, then what matters is that the causal
horizon can encompass a fluxon.

Although some caution is required in the interpretation of past
experiments\cite{monaco} with JTJs, for which ${\bar\xi}\approx
C$, which were not devised with this prediction in mind, we would
have expected to see a fluxon a few percent of the time if the
latter were true. This was the case. If the former mechanism is
correct, we would have expected to see them more often. New
experiments with more relevant parameters are being devised and
have been reported elsewhere\cite{KMR,KMR2}.

In a different context an experiment with an annular
superconductor, separated into segments by JTJs, shows that the
assumed phase separation at the onset of a transition (the 'Kibble
mechanism' in the context of QFT) is correct\cite{polturak}. A
further experiment, to quench an annular container of similar
circumference $C$ of superfluid $^{4}He$, was also proposed by
Zurek\cite{zurek1}, but has not been attempted. Since the phase
gradient is directly proportional to the superflow velocity we
expect a flow after the quench with r.m.s velocity
\begin{equation}
\Delta v =
O\bigg(\frac{\hbar}{m}\sqrt{\frac{1}{C{\bar\xi}}}\bigg).
\label{v1}
\end{equation}
Although not large it is measurable, in principle. A similar
experiment for BEC has been proposed\cite{anglin}, but has also
not been performed.

\section{CONCLUSIONS}

The strong causal bounds (\ref{bart}) and (\ref{zcaus}) of Zurek
and (\ref{bartk}) and (\ref{kcaus}) of Kibble have a common origin
that purports to show strong links between phase transitions in
condensed matter and particle physics. In reality, this masks huge
differences between the two types of system. Nonetheless, each has
a certain qualitative validity, but for different reasons. What we
have found in common is that explicit causality is absent from
both.

In TDGL condensed matter physics we have seen how fluctuations
separate initially into unstable long wavelength modes which order
the field and short wavelength thermal fluctuations that make
vortices (and other defects) fractal. Dimensional analysis sets
the Zurek time ${\bar t}_Z$ of (\ref{bart}) and length scale
${\bar\xi}_Z$ of (\ref{zcaus}) as the natural scales when
fluctuations can be ignored. However, the density of defects is
not set by ${\bar\xi}_Z$ (as the correlation length of the field)
since it freezes too soon for  instabilities to have grown.
Instead, the density is set by the separation of field zeroes, a
very different length, in principle.

In practice, because instabilities grow exponentially, the time it
takes to order the field is, qualitatively, $O({\bar t}_Z)$, with
only logarithmic dependence on the parameters of the system.
Further, since the defect density is, essentially, a ratio of
moments of the power spectrum of the fluctuations, there is strong
cancellation of the prefactors that contain the detailed
information of the system. As a result, if thermal fluctuations
are negligible the defect density is that of a free field for
longer than we might have thought, and Zurek's causal predictions
are valid, although not for directly causal reasons. If we want to
see explicit causality in the TDGL formalism we need to go to its
dual worldline representation, which we have not discussed here,
but which we have examined elsewhere\cite{rr}.

Although thermal fluctuations and long wavelength (unstable) modes
are not additive in determining defect densities, simple analysis
suggests that thermal fluctuations are roughly proportional to the
current temperature, with little memory of the past temperature
profile. As a result, the Ginzburg regime plays no role in
incubating defects.

Quantum Field Theory has some characteristics in common with
condensed matter: exponential (albeit different exponential)
behaviour  makes ${\bar t}_K$ the natural time-scale, up to
logarithmic terms. We recover ${\bar\xi}_K$ as the relevant
length-scale after instabilities have formed, rather than before.
However, there is a potential complication in that we can only
adopt a classical probabilistic approach to defects once there are
no substantial quantum interference effects. We have shown that
decoherence is established before field ordering in a simplified
model and more work is being done. However, once the initial
defects are established, qualitatively as suggested by Kibble,
their long-term behaviour is unclear. Dissipation is determined by
Landau damping and long wavelength noise is now multiplicative.
Unlike the case of the TDGL equation, where we have many numerical
calculations, we have very little for QFT. As before, the Ginzburg
regime only looks to make defects fuzzy, rather than create them.

As for prediction and experiment, we have no concrete proposals
for QFT. For condensed matter, experiments in superfluid $^3He$
are in agreement with Zurek-Kibble, while those for $^4He$ seem
not to be. However, we understand this as due, in large part, to
an over-optimistic assumption about their decay rate. The role of
large thermal fluctuations is unclear. Although the one experiment
with High-$T_c$ superconductors sees nothing, an earlier
experiment with Josephson Tunnel Junctions seems to give support
to Zurek's predictions, and further experiments with JTJs are
planned.

In none of these experiments is there any agreement with the
predictions of thermal production in the Ginzburg regime.

\section*{ACKNOWLEDGEMENTS}
I thank Roberto Monaco, Fernando Lombardo, Diego Mazzitelli and
Eleftheria Kavoussanaki for enjoyable collaborations. This work is
supported by the European Science Foundation.

\end{document}